\documentclass[fleqn,usenatbib]{mnras}

\usepackage{newtxtext,newtxmath}

\usepackage[T1]{fontenc}

\DeclareRobustCommand{\VAN}[3]{#2}
\let\VANthebibliography\thebibliography
\def\thebibliography{\DeclareRobustCommand{\VAN}[3]{##3}\VANthebibliography}

\usepackage{graphicx}	
\usepackage{amsmath}	
\usepackage{threeparttable} 
\usepackage{url}
\usepackage{subcaption}
\usepackage{float}
\usepackage{xcolor}
\usepackage{upgreek}
\usepackage{caption}
\DeclareCaptionFont{magenta}{\color{magenta}}

\usepackage{tikz,xcolor,hyperref}

\definecolor{lime}{HTML}{A6CE39}
\DeclareRobustCommand{\orcidicon}{
    \begin{tikzpicture}
    \draw[lime, fill=lime] (0,0)
    circle [radius=0.16]
    node[white] {{\fontfamily{qag}\selectfont \tiny ID}};
    \draw[white, fill=white] (-0.0625, 0.095)
    circle [radius=0.007];
    \end{tikzpicture}
    \hspace{-2mm}
}

\foreach \x in {A, ..., Z}{\expandafter\xdef\csname orcid\x\endcsname{\noexpand\href{https://orcid.org/\csname orcidauthor\x\endcsname}
			{\noexpand\orcidicon}}
}

\newcommand{\snr}{\,G\,116.6$-$26.1\,} 
\newcommand{\Hafilt}{H$\upalpha$+$[\ion{N}{ii}]$\,} 
\newcommand{\SIIfilt}{$[\ion{S}{ii}]$\,} 
\newcommand{\OIIIfilt}{$[\ion{O}{iii}]$\,} 
\newcommand{\ha}{H$\upalpha$\,} 
\newcommand{\hb}{H$\upbeta$\,} 
\newcommand{\nii}{$[\ion{N}{ii}]$\,}
\newcommand{\fluxu}{$10^{-18}$\,erg\,\AA$^{-1}$\,s$^{-1}$\,cm$^{-2}$\,arcsec$^{-2}$}

\title[Optical identification of SNR \snr]{First optical identification of the
\textit{SRG}/eROSITA-detected supernova remnant \snr\; I. Preliminary results.}

\author[E.V. Palaiologou et al.]{
E.V. Palaiologou,$^{1,2}$\thanks{e-mail: palaiolo@physics.uoc.gr} I. Leonidaki,$^{1,2\orcidB}$\ M. Kopsacheili$^{1,2\orcidC}$
\\
$^{1}$University of Crete, Department of Physics, Voutes University Campus, GR-70013 Heraklion, Greece\\
$^{2}$Foundation for Research and Technology -- Hellas, P.O. Box 1385, GR-71110 Heraklion, Greece
}

\date{Accepted 2022 June 02. Received 2022 June 01; in original form 2022 April 20}
\pubyear{2022}

\begin{document}
\label{firstpage}
\pagerange{\pageref{firstpage}--\pageref{lastpage}}
\maketitle

\begin{abstract}
The supernova remnant (SNR) candidate \snr is one of the few high Galactic latitude (|b|\,>\,15$^\circ$)
remnants detected so far in several wavebands. It was discovered recently in the \textit{SRG}/eROSITA
all-sky X-ray survey and displays also a low-frequency weak radio signature. In this study,
we report the first optical detection of \snr through deep, wide-field and higher resolution
narrowband imaging in \ha, \SIIfilt and \OIIIfilt light. The object exhibits two major and distinct
filamentary emission structures in a partial shell-like formation. The optical filaments are
found in excellent positional match with available X-ray, radio and UV maps, can be traced over
a relatively long angular distance ($38\arcmin$ and $70\arcmin$) and appear unaffected by any strong
interactions with the ambient interstellar medium. We also present a flux-calibrated,
optical emission spectrum from a single location, with Balmer and several forbidden lines detected,
indicative of emission from shock excitation in a typical evolved SNR. Confirmation of the most
likely SNR nature of \snr is provided from the observed value of the line ratio 
[\ion{S}{ii}]\,/ \ha = $0.56\,\pm \,0.06$, which exceeds the widely accepted threshold 0.4,
and is further strengthened by the positive outcome of several diagnostic tests for shock emission.
Our results indicate an approximate shock velocity range 70--100 km s$^{-1}$ at the spectroscopically
examined filament, which, when combined with the low emissivity in \ha and other emission lines,
suggest that \snr is a SNR at a mature evolutionary stage.

\end{abstract}




\section{Introduction} \label{sec:intro}

\snr is a newly discovered Galactic supernova remnant (SNR) through soft X-ray imaging
and spectra obtained with the Russian-German observatory \textit{SRG}/eROSITA \citep{Churazov_2021}.
According to the X-ray observations, it is a faint and nearly circular object, with large angular size
(about 4$^{\circ}$ in diameter), located in high Galactic latitude.
The morphological structure of the SNR candidate indicates a marginal brightening along the periphery
(mostly in the southern and western portions) and signs of mild surface brightness enhancement at the
innermost 20\arcmin of the object. Its soft X-ray imaging spectrum is dominated by \ion{O}{vii} and 
\ion{O}{VIII} lines, similar to the surrounding background spectrum.
\citet{Churazov_2021} suggest that the supernova (SN) explosion took place in the halo of Milky Way,
characterized by hot [$\sim$(1--2)\,$\times\,10^6\,K$] and low density (about $10^{-3}~cm^{-3}$) plasma. 

They propose a relatively old remnant, originating from a thermonuclear explosion (SN Ia)
that happened  about 40\,000 years ago, at an approximate distance of 3~kpc from us
and about 1.3~kpc out of the Galactic disk. However, they do not reject the possibility of \snr being
the result of a core collapse SN explosion. In favor of this scenario is the fact that probably the
dust distribution in its neighborhood is affected by \snr, which would imply that this object
is placed within 300~pc from us, resulting from a core collapse supernova explosion (SN II). 
In a very recent work, \citet{Churazov_2022} report radio emission from \snr detected in the
LOFAR Two-metre Sky Survey \citep[LoTTS-DR2,][]{Shimwell_2022}. It presents a faint,
shell-like structure with good positional coincidence between the radio boundary and the X-ray limb,
as presented in \cite{Churazov_2021}, while no optical counterparts have been reported until now.

In our Galaxy there are roughly 300 known SNRs \citep{Green_2019}, with more than 90 per cent of them
located within 5 degrees off the Galactic plane \citep{Kothes_2017},
while $\sim$\,10 of the currently known remnants reside between latitudes 5 and 10 degrees.
However, only a handful of these objects have been found in higher Galactic latitudes.
\citet{Boumis_2002} reported the discovery of optical filaments in Pegasus Constellation
suggesting that are part of one (or more) SNRs. Its nature as such was also confirmed by
\citet{Fesen_2015} based on its morphology and its spectral characteristics.
This is the G\,70.0$-$21.5 remnant which was studied more recently also by \citet{Raymond_2020},
and it is believed to be the result of a type Ia SN.
\citet{McCullough_2002} discovered the Antlia SNR (or G\,275.5$+$18.4),
which was detected in \ha and X-rays. Later, \citet{Shinn_2007} and  \citet{Fesen_2021} confirmed
this identification, using GALEX FUV imagery and optical observations, respectively.
Its progenitor is probably a B-type star \citep{Shinn_2007} and hence, it comes likely
from a core collapse SN explosion.

Another high latitude SNR is the Hoinga or G\,249.7$+$24.7 remnant discovered also in the X-ray
\textit{SRG}/eROSITA All-Sky Survey eRASS1 \citep{Becker_2021}, which
presents also radio, optical, and UV emission \citep{Becker_2021, Fesen_2021}.
According to \citet{Becker_2021}, the fact that no pulsar has been associated with this remnant so far,
in combination with its high latitude, indicate that it is probably the remnant of
a type Ia SN explosion.
The  highest Galactic latitude SNR found yet, is G\,354.0$-$33.5,
observed in  FUV, \ha, and radio wavelengths \citep{Testori_2008, Fesen_2021}.
No evidence about its progenitor and/or the explosion type of this SNR is found.

In this work we report for the first time optical emission from \snr, based on both wide-field
and higher resolution imaging as well as spectroscopic observations. The results seem to satisfy most
of the criteria for the optical identification of SNRs. This, in combination with its spatial
correlation with other wavelengths, enhances the belief of \snr being a new Galactic SNR.
We also discuss the possible nature of its progenitor but we did not reach a secure conclusion.
The structure of the paper is as follows:
In Section \ref{sec:observations} we describe the observations and the reduction of the imaging and
spectroscopic data. In Section \ref{sec:results} we present the results of the aforementioned analysis.
We quote available observations of \snr in other wavelengths and we explore their spatial correlation
with the optical emission in Section \ref{sec:other_wavebands}. In Section \ref{sec:discussion}
we discuss our results, while in Section \ref{conlcusions} we summarise our conclusions.

\section{Observations and Data Reduction} \label{sec:observations}

\subsection{Imaging}
\label{sec:imaging}

\subsubsection{Wide-field imagery}
The first step in our search for possible optical line emission from the newly-discovered SNR \snr,
was to perform wide-field imaging, since the reported angular diameter of the source in X-rays was
estimated to be ${\sim}4\degr$ \citep{Churazov_2021}. Such a capability is provided by the fast-optics
(f/3.2), Schmidt-Cassegrain 0.3~m telescope (a Lichtenknecker Flat Field Camera, hereafter LFFC) 
at Skinakas Observatory in Crete, Greece. Coupled with a back-illuminated 2k $\times$ 2k CCD camera
(by Andor Tech.), the telescope--instrument system offers a field of view (FoV) $100\arcmin \times
100\arcmin$ on the sky. The thermoelectric-cooled CCD sensor (operating temperature $-70\degr$ Celsius)
shows negligible dark current, has pixel size 13.5~$\micron$ which corresponds to $2\farcs94$ on the
celestial sphere.

This preliminary observing run aimed at the construction of a mosaic of partially overlapping images,
wide enough to cover the reported X-ray extent of the source, in the light of \ha emission,
in a $3\times3$ frame pattern. The images were obtained with the LFFC telescope
on 2021 September 1, 3--5 and October 7--8, through a narrow-band \Hafilt filter.
A red broadband continuum filter (SDSS$-$r') was also used to provide images for
subsequent subtraction of the background continuum and especially for the removal of the stellar
component in the images in order to improve the visibility of very faint structures.
Filter characteristics are given in Table~\ref{tab:filter_chars}. Each of the nine fields
in the mosaic was observed for a total exposure time of 9000~s (15$\times$600~s) in \Hafilt\
and 300~s (15$\times$20~s) through the continuum filter, respectively.

All nights were photometric and target fields ranged in airmass between 1.0 and 1.4 during observations. 
Multiple bias exposures and twilight flat frames of very high signal-to-noise (S/N) ratio
were taken on a daily basis. We did not observe any spectrophotometric standard stars,
because we were not planning to flux-calibrate our images taken with the LFFC 0.3-m telescope,
serving the single purpose to visually examine the area for any H$\upalpha$ emission signature.

\begin{table}
	\centering
	\caption{The characteristics of the optical filters used in this research.}
	\label{tab:filter_chars}
	\begin{threeparttable}[b]
		\begin{tabular}{lccc} 
			\hline
			Filter & $\uplambda_{c}$\tnote{a} & FWHM & Transm\tnote{b}\\
			& (\AA) & (\AA) & (\%)  \\
			\hline
			\Hafilt~$\uplambda\uplambda$ 6548, 6584~\AA & 6582 & 82 & 99 \\
			\SIIfilt~$\uplambda\uplambda$ 6716, 6731~\AA & 6730 & 32 & 99\\ 
			\OIIIfilt~$\uplambda$ 5007~\AA & 5008 & 25 & 52\\
			\,SDSS$-$r' & 6214 & 1290 & 96 \\
			\,Johnson/Bessell V & 5380 & 980 & 88 \\
			\hline
		\end{tabular}
		\begin{tablenotes}
			\item[a] The central wavelength
			\item[b] Peak filter transmittance
		\end{tablenotes}
	\end{threeparttable}
\end{table}

\begin{figure*}
	\includegraphics[width=0.95\textwidth]{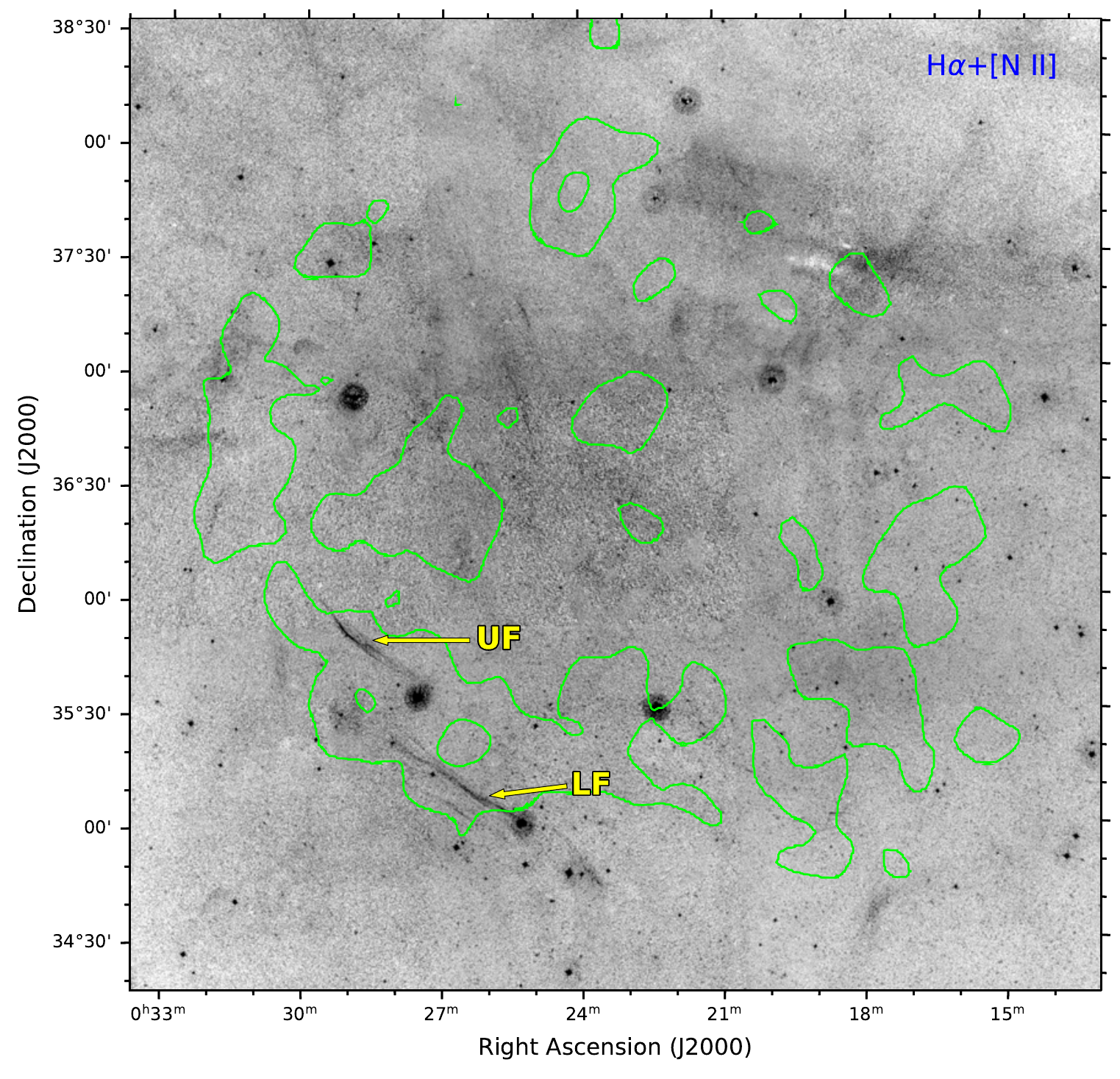}
	\caption{The continuum-subtracted 9-frame mosaic towards the SNR \snr in the light of \Hafilt.
		Superposed contours (solid green lines) from the X-ray observations \citep{Churazov_2021}
		show the correlation between the optical filaments with the X-ray emission.
		Several artefacts in the form of dark round halo appear around bright stars
		which are not emission features but ghosts due to scattered light from internal reflections
		in the narrow-band filter.}
	\label{fig:Ha_mosaic_30cm}
\end{figure*}

The raw images were reduced using the \textsc{IRAF} software \citep{iraf_1986, iraf_1993} 
procedures for bias subtraction and flat-field correction. Individual exposures of each
of the 9 frames observed were registered to a reference image in the set and combined
through a 3-sigma clipping average algorithm in order to remove cosmic rays and artificial
satellite trails -- a not so rare situation in modern era, especially in wide-field imaging.
Astrometric calibration was performed with the aid of
\texttt{astrometry.net} \citep{astrometry.net} web service,
interfaced through \textsc{AstroImageJ} software \citep{Collins_2017}. The next step was to use
the narrow-band (\Hafilt) and continuum (SDSS-r') images in order to subtract the sky background
and eliminate as many field stars as possible to reduce crowding confusion, following a procedure
described in Appendix \ref{appendix:image_subtraction}. Finally, the continuum-subtracted
frames were assembled together into the final mosaic utilizing the
\textsc{Montage}\footnote{\url{http://montage.ipac.caltech.edu}} software with adjustment for
smooth background levels between overlapping images. The final mosaic in \Hafilt emission is shown
in Fig.~\ref{fig:Ha_mosaic_30cm}, where the presence of two major optical filaments is evident.

\begin{figure}
	\centering
	\begin{subfigure}{\columnwidth}
		\centering
		\includegraphics[width=0.90\columnwidth]{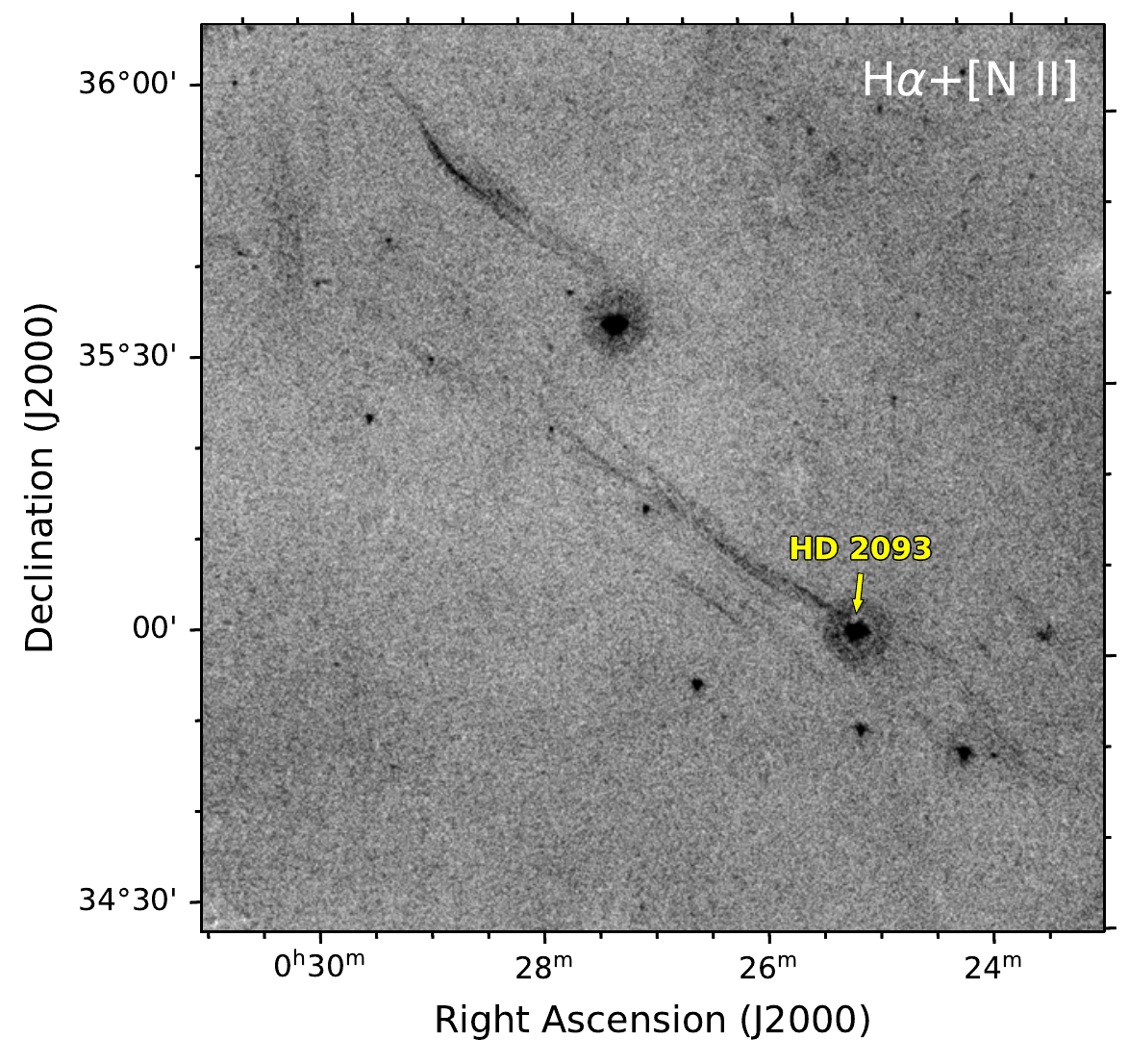}
		\label{fig:Flm_30cm_Ha}
	\end{subfigure}
	\vspace{-6pt}
	\begin{subfigure}{\columnwidth}
		\centering
		\includegraphics[width=0.90\columnwidth]{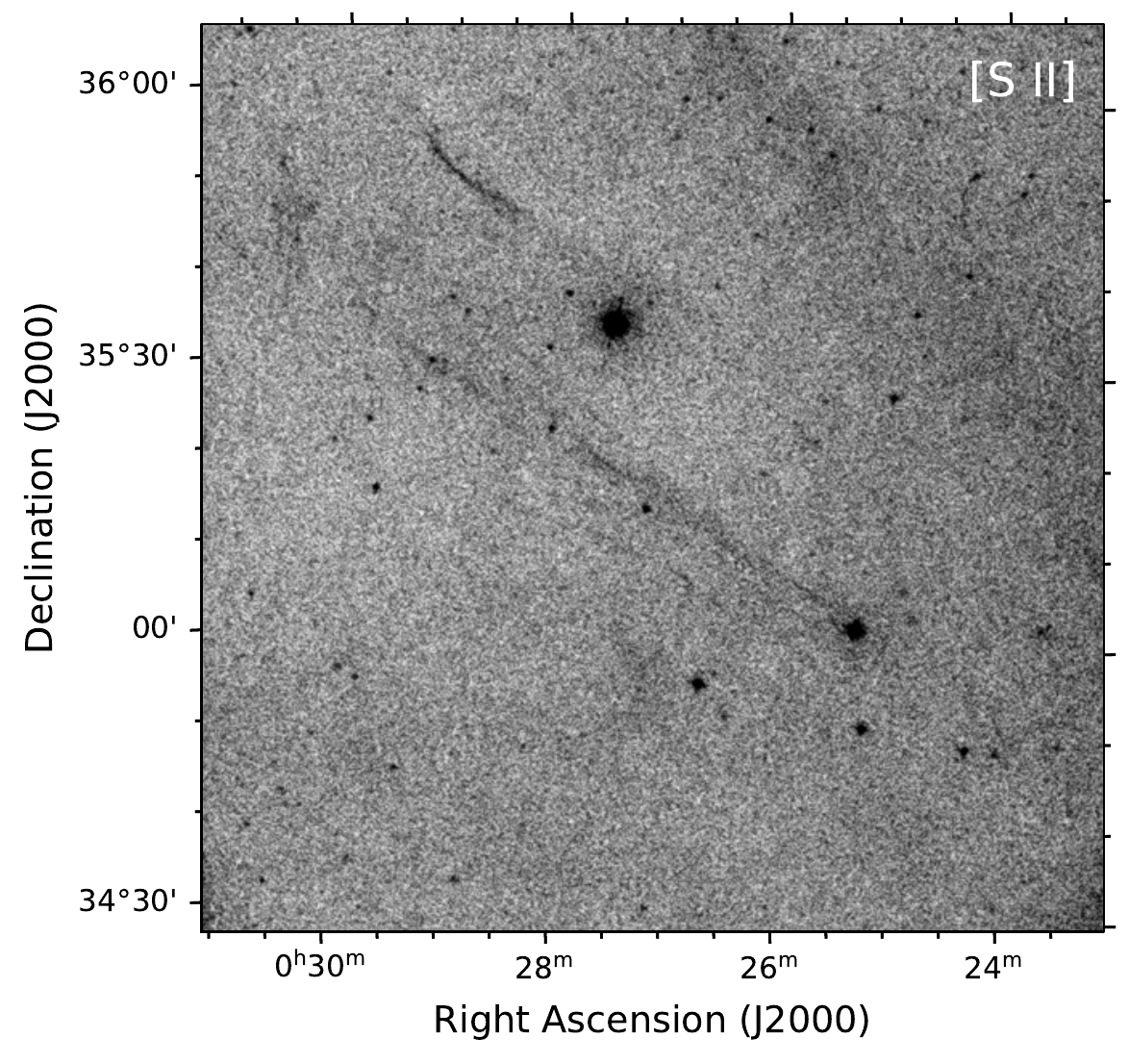}
		\label{fig:Flm_30cm_SII}
	\end{subfigure}
	\vspace{-5pt}
	\begin{subfigure}{\columnwidth}
		\centering
		\includegraphics[width=0.90\columnwidth]{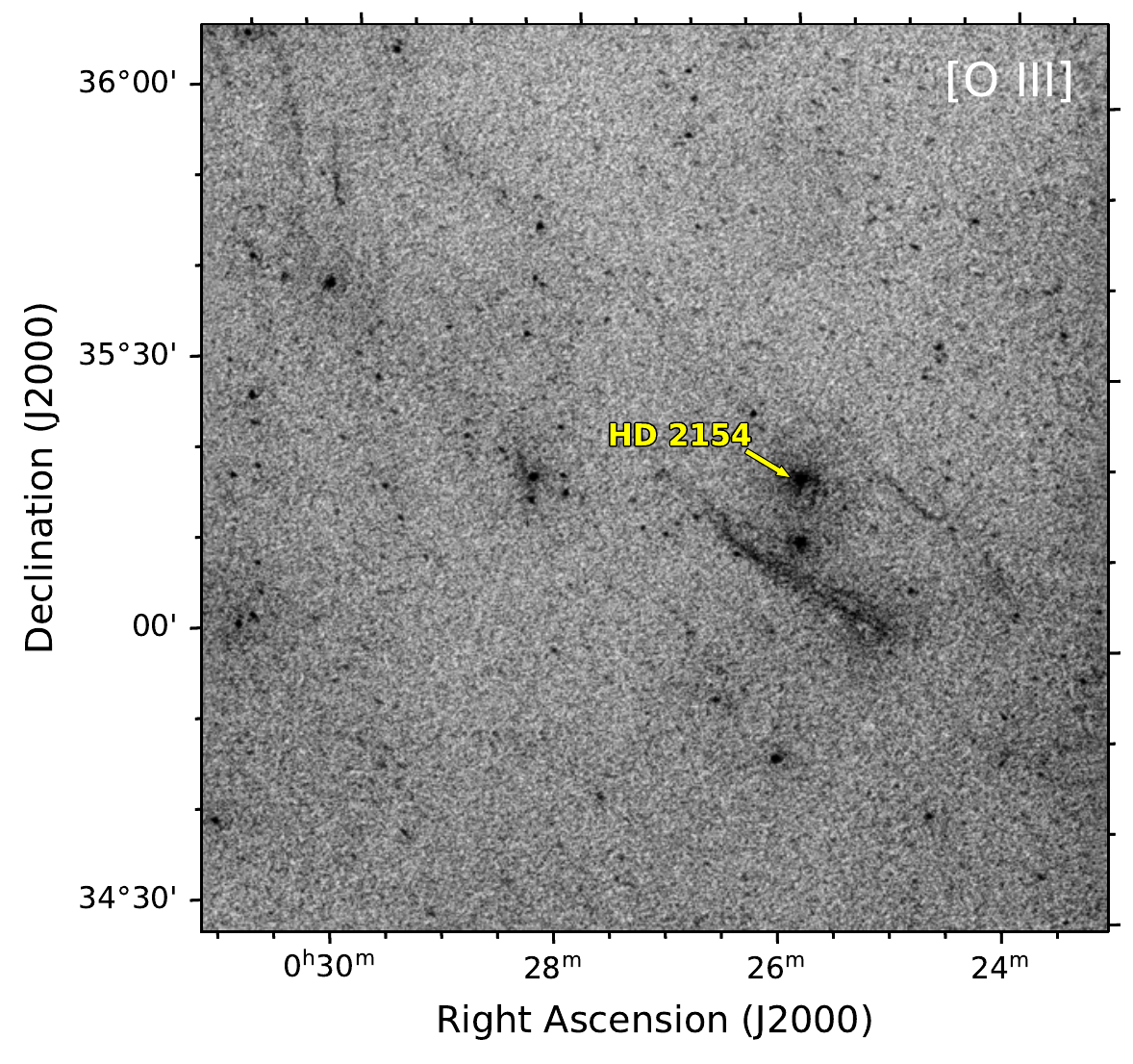}
		\label{fig:Flm_30cm_OIII}
	\end{subfigure}
	\vspace{-5pt}
	\caption{Low resolution negative images of the major optical filaments' area, in
	\Hafilt\, (top), \SIIfilt\,
		(middle) and \OIIIfilt\, (bottom panel). Note that the dark spots and disc-like features
		scattered around the images, are residuals from	imperfect subtraction of bright stars.
		These leftover artefacts in the bottom panel (\OIIIfilt\, filter) have no counterpart
		in the top and middle (red filters) images.}
	\label{fig:Flm_30cm}
\end{figure}

The detected sharp filaments shown in Fig.~\ref{fig:Ha_mosaic_30cm} were further investigated, 
using the LFFC telescope and \Hafilt, \SIIfilt\ and \OIIIfilt filters, on 2021 November 6 and 8,
for a total exposure time of 3~h (18 images of 600~s each), 4~h (24$\times$600~s)
and 3~h (9$\times$1200~s), through the respective filters. Multiple short exposures were also
obtained in the respective continuum filters for background and starlight subtraction.
The final background-subtracted images are shown in Fig.~\ref{fig:Flm_30cm}.

\begin{figure}
	\centering
	\includegraphics[width=\columnwidth]{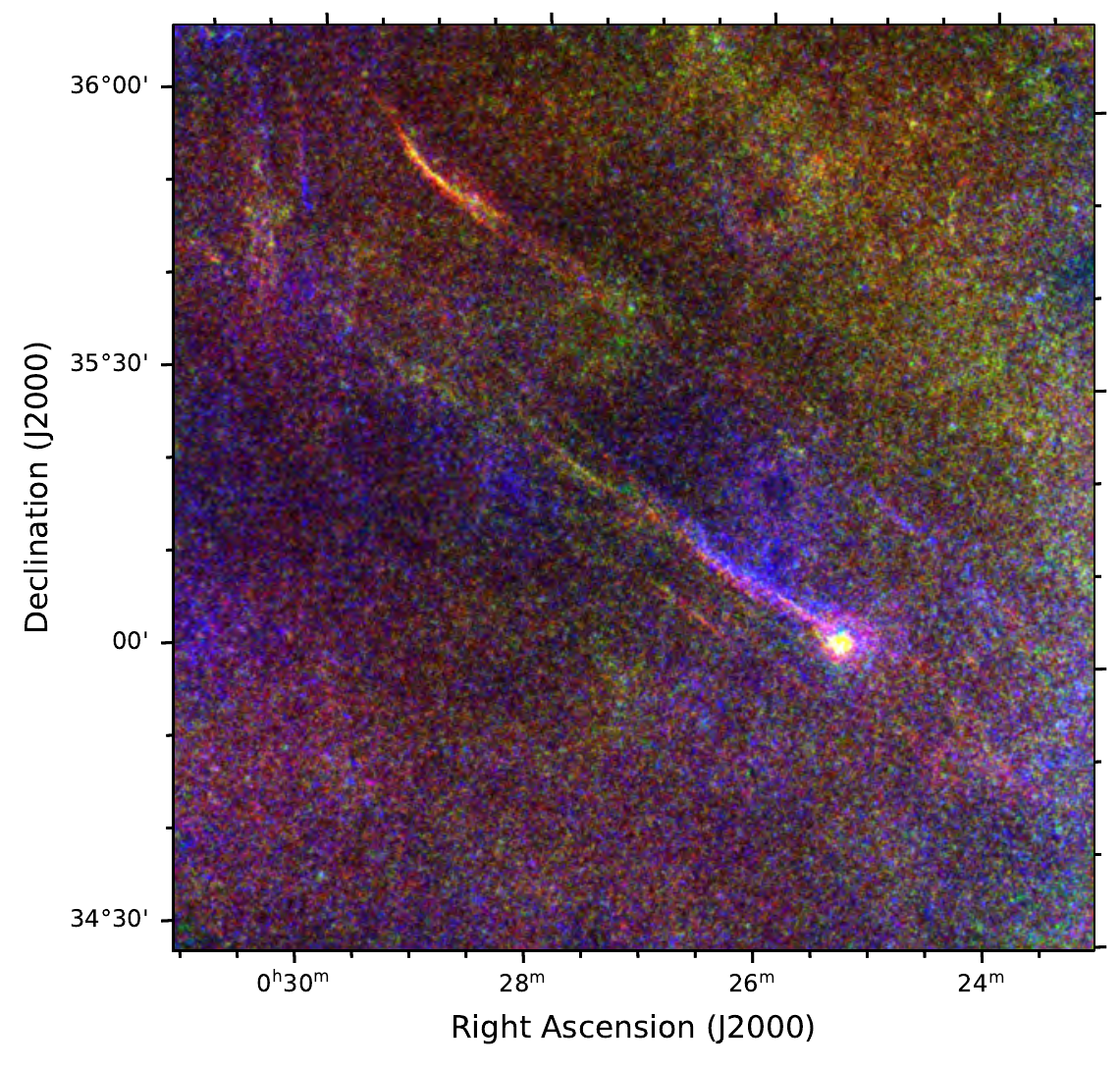}
	\caption{The area of the major optical filaments (Fig.~\ref{fig:Flm_30cm}) in \Hafilt (red),
	\SIIfilt (green) and \OIIIfilt (blue), to indicate the relative position of the emission
	in these three lines near the shock fronts.}
	\label{fig:Flm_30cm_color}
\end{figure}

\subsubsection{Higher angular resolution imaging} \label{sec:hi_res_imaging}
Guided by the wide-field mosaic of the object's \Hafilt\ emission, additional
images were taken of the upper and lower optical filaments at the south-eastern edge of the
X-ray structure, using the 1.29~m (f/7.6) Ritchey--Chr\'{e}tien telescope at Skinakas Observatory.
Images were acquired with a 2048 $\times$ 2048 back-illuminated, deep-depletion CCD camera
(Andor iKon-L) with pixel size 13.5~$\micron$. The system provided a 9$\farcm$6 $\times9\farcm6$
FoV and image scale $0\farcs28$ per pixel.
The narrow-band filters used are these in Table~\ref{tab:filter_chars}.
Observations were held on 2021 July 15, September 10 and 12, October 4--5 and
November 5, 7 and 11, under good seeing conditions ($0\farcs8$--$1\farcs2$) and high target elevation
(airmass range 1.00--1.35). Images for a 3-frame mosaic were obtained in the \Hafilt\ filter covering
the brighter part of the upper filament (region UF in Fig.~\ref{fig:Ha_mosaic_30cm})
with 9600~s (16$\times$600~s) total integration time for each frame.
Another series of 16 $\times$ 600~s exposures in \SIIfilt combined into a single-frame image of
the brighter part of UF. The lower filament (designated region LF in Fig.~\ref{fig:Ha_mosaic_30cm})
was also imaged -- only with a partial filter coverage due to weather conditions -- in \Hafilt\
with 2 overlapping frames of 2~h total exposure time each ($12 \times 600$~s).

\begin{table*}
	\centering
	\caption{Spectroscopic observations log.}
	\label{tab:spectra_info}
	\begin{threeparttable}[b]
		\begin{tabular}{lcccccc} 
			\hline
			RegionID & Date & \multicolumn{2}{c}{Aperture center} & Aperture & Exposures\tnote{a} &
			$\mathrm{X_{eff}}$\tnote{b} \\
			& & R.A. (J2000) & Dec (J2000) & height &  & \\
			&(UT)&  & & (\arcsec) & (s) & \\
			\hline
			UF & 2021 Nov 3 & $\mathrm{00^{h}29^{m}03\fs7}$ & $\mathrm{+35^{d}51^{m}50^{s}}$ & 22.5 &
			$3\times2400$ & 1.035,\,1.008,\,1.002 \\ 
			\hline
		\end{tabular}
		\begin{tablenotes}
			\item[a] Number of individual exposures and common exposure time 
			\item[b] Effective airmass at each exposure.
			[$\mathrm{X_{eff}~=~\frac{1}{6}(X_{start}\,+\,4\,X_{mid}\,+\,X_{end})}$]
		\end{tablenotes}
	\end{threeparttable}
\end{table*}

\begin{table}
	\centering
	\captionsetup{labelfont={color=magenta},font={color=magenta}}
	\caption{Locations of fainter filaments detected in wide-field images, but not examined yet in detail.}
	\label{tab:other_flms}
	\begin{threeparttable}[b]
		\begingroup
		\setlength{\tabcolsep}{4pt} 
		\begin{tabular}{lcccc} 
			\hline
			\multicolumn{2}{c}{Filament center} & Approximate & Appears & Image \\
			R.A. (J2000) & Dec (J2000) & length (\arcmin) & in figure & filter \\
			\hline
			$\mathrm{00^{h}25^{m}15^{s}}$ & $\mathrm{+37^{d}15^{m}28^{s}}$ & 6.7 &
			Fig.~\ref{fig:Ha_mosaic_30cm} & \Hafilt \\
			$\mathrm{00^{h}25^{m}22^{s}}$ & $\mathrm{+36^{d}52^{m}40^{s}}$ & 18.5 &
			Fig.~\ref{fig:Ha_mosaic_30cm} & \Hafilt \\
			$\mathrm{00^{h}24^{m}22^{s}}$ & $\mathrm{+34^{d}57^{m}33^{s}}$ & 7.7 &
			Fig.~\ref{fig:Ha_mosaic_30cm} & \Hafilt \\
			$\mathrm{00^{h}24^{m}46^{s}}$ & $\mathrm{+34^{d}55^{m}09^{s}}$ & 6.9 &
			Fig.~\ref{fig:Ha_mosaic_30cm} & \Hafilt \\
			$\mathrm{00^{h}24^{m}08^{s}}$ & $\mathrm{+35^{d}01^{m}07^{s}}$ & 2.6 &
			Fig.~\ref{fig:Ha_mosaic_30cm} & \Hafilt \\
			$\mathrm{00^{h}30^{m}10^{s}}$ & $\mathrm{+35^{d}51^{m}28^{s}}$ & 8.6 &
			Fig.~\ref{fig:Flm_30cm} & \OIIIfilt \\
			$\mathrm{00^{h}30^{m}40^{s}}$ & $\mathrm{+35^{d}58^{m}48^{s}}$ & 13.0 &
			Fig.~\ref{fig:Flm_30cm} & \OIIIfilt \\
			$\mathrm{00^{h}25^{m}00^{s}}$ & $\mathrm{+35^{d}18^{m}13^{s}}$ & 12.6 &
			Fig.~\ref{fig:Flm_30cm} & \OIIIfilt \\
			$\mathrm{00^{h}31^{m}38^{s}}$ & $\mathrm{+37^{d}08^{m}18^{s}}$ & 21.3 &
			Fig.~\ref{fig:GALEX_FUV}  & Far UV \\
			\hline
		\end{tabular}
	\endgroup
	\end{threeparttable}
\end{table}

\subsection{Spectrophotometry} \label{sec:spectrophoto}
Long-slit, low-resolution spectroscopic observations of the brightest part of the upper filament
were made on 2021 November 3, using the 1.29~m telescope at Skinakas Observatory
and a CCD camera of the same type and characteristics as in the higher resolution direct imaging,
described in section \ref{sec:hi_res_imaging}.
A 1302 grooves mm$^{-1}$ grating was used, blazed at 5500 \AA, giving an average dispersion
0.94~\AA~pixel$^{-1}$ and spectral coverage 4820--6750~\AA. The object spectra were taken with
a slit of width 3\farcs9 (7\farcs3 slit used for the spectrophotometric standard stars), and usable
length 14\farcm5 -- long enough to allow for background subtraction -- always oriented
along the North-South direction. Seeing varied between 1\farcs1 and 1\farcs5 under dark and photometric
sky conditions.

Several zero-exposure (bias) images were collected each night along with two sets of exposures for
flat-fielding. One set consisted of images taken with a quartz-lamp as light source inside the
spectrograph and a diffuser in front of the slit, while the second set was a series of twilight
sky exposures, in order to perform a slit illumination correction on the first set.
Spectra were wavelength calibrated using a Ferrum-Helium-Neon-Argon comparison lamp and arc images
were obtained after the end of each target exposure.
Flux calibration of the spectral lines was achieved through exposures of the spectrophotometric
standard stars HR718, HR1544, HR3454, HR7596, HR7950, HR9087 \citep{Hamuy_1992, Hamuy_1994}
and BD$+25{\degr}4655$, BD$+33 \degr 2642$, G\,191-B2B \citep{Oke_1990, Massey_1988}.
Standards were observed every night in at least three different time slots,
in groups of 4--6 stars for better airmass coverage.

One more set of images was acquired, necessary for the correction of optical distortion
and curvature of the spectral features along the dispersion and spatial directions, which is
quite noticeable in our long-slit spectra. A preselected nearby (angular distance \la~8\degr\,
from the observed target), very bright field star (visual magnitude 3.5--5.5) was positioned near one
of the slit's long-dimension ends and a series of spectral images was obtained, offseting the
telescope in steps of $\approx 30\arcsec$ along the slit direction in-between exposures,
until the other end of the slit was reached. This task is accomplished in less than 5 minutes,
since typical exposure times for high signal-to-noise ratio individual spectra is 3--5~s.
The images are added together during data reduction and give the equivalent of a multi-star spectrum,
which, along with a comparison lamp exposure, can determine the geometric transformation needed
to correct distortions in the two-dimensional science spectra.

Standard \textsc{IRAF} procedures \citep{Valdes_1992} were utilised for spectra reduction and
extraction of calibrated line emission fluxes. The reduction steps include bias subtraction,
flat-field correction, cosmic-ray removal using the software \textsc{L. A. Cosmic} \citep{vanDokkum_2001}
and geometric rectification \footnote{\url{https://iraf.net/irafdocs/spect.pdf}}.
Background estimation was performed in areas close to the spectral apertures and free
of diffuse emission, selected with the guidance of the higher resolution images.
Center coordinates and height of the aperture used in the spectra extraction,
exposure date, integration times and effective airmasses are presented in Table~\ref{tab:spectra_info}.
\begin{figure*}
	\begin{minipage}[b]{0.66\textwidth}
		\includegraphics[width=\linewidth]{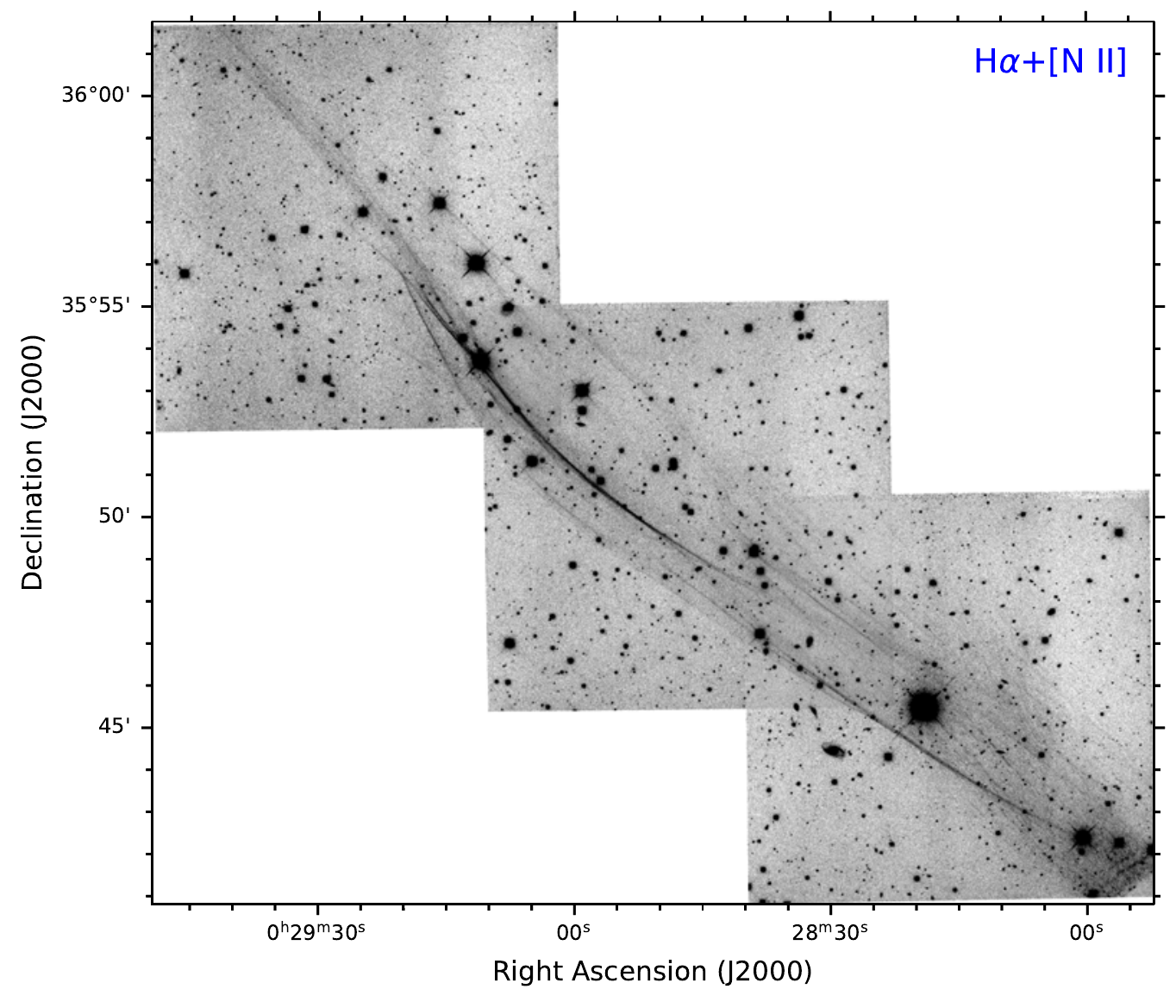}
		\caption{Higher angular resolution mosaic in \Hafilt\, of the upper filament's (UF)
			brighter segment. An orderly network of similarly aligned, hair-thin filaments of varying
			degree of curvature is prominent in this image.}
		\label{fig:UF_129cm_Ha} 
	\end{minipage}
	\hfill
	\raisebox{151pt}
	{
		\begin{minipage}[c]{0.307\textwidth}
			\includegraphics[width=\linewidth]{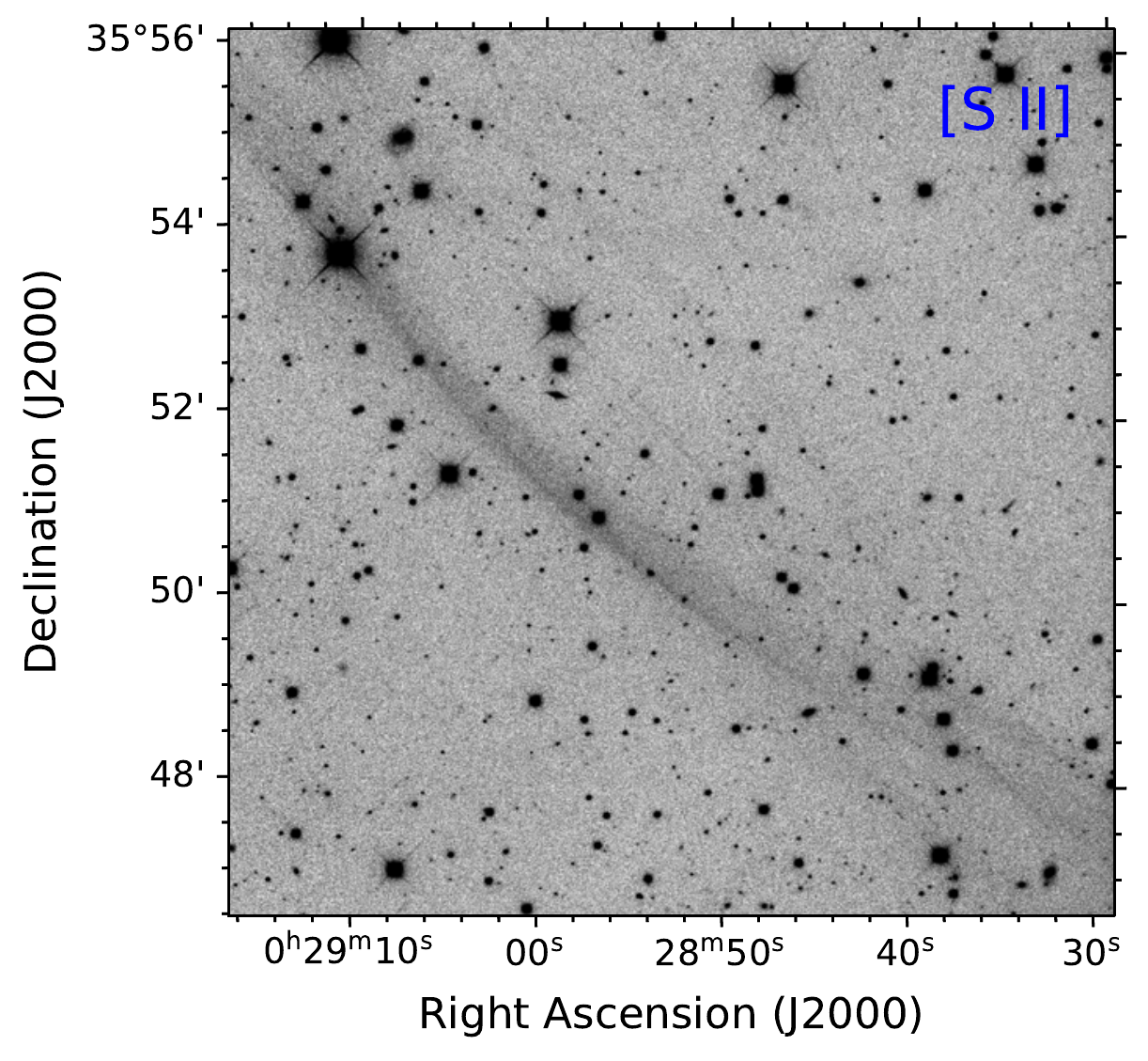}
			\caption{The middle frame of the \Hafilt\, mosaic in Fig.~\ref{fig:UF_129cm_Ha},
				as seen in \SIIfilt\, emission light. Although images in both filters were obtained under
				the same seeing conditions, the filaments' sharpness in \SIIfilt\, is not comparable
				to that in \ha light. }
			\label{fig:UF_129cm_SII}
		\end{minipage}
	}
\end{figure*}

\begin{figure*}
	\begin{minipage}[t]{0.45\textwidth}
	\includegraphics[width=\textwidth]{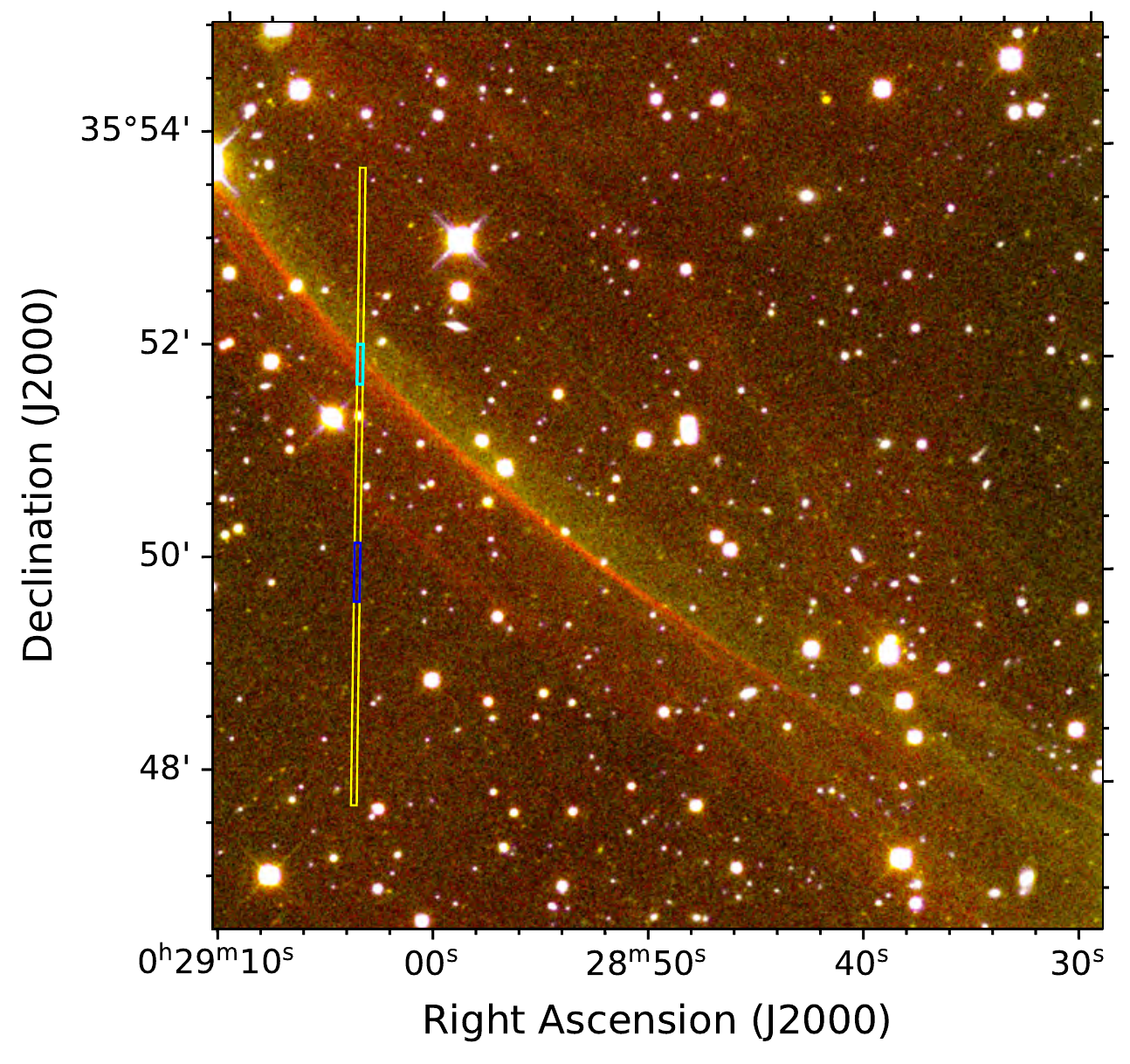}
	\caption{Uncalibrated bi-colour composite image of the area in Fig.~\ref{fig:UF_129cm_SII} with
	\Hafilt\, (red) and \SIIfilt\, (green). The respective emissions at the brightest filament area
	seem to start at the same location (dark orange coloured post-shock shell), but the
	\ha\, intensity drops faster in the downstream direction compared to \SIIfilt.
	The yellow, cyan and blue rectangles mark the locations of the slit, aperture for the extracted
	spectrum and sky area for background subtraction, respectively.}
	\label{fig:UF_129cm_HaSII}
\end{minipage}
\hfill
{
\begin{minipage}[t]{0.53\textwidth}
	\includegraphics[width=\textwidth]{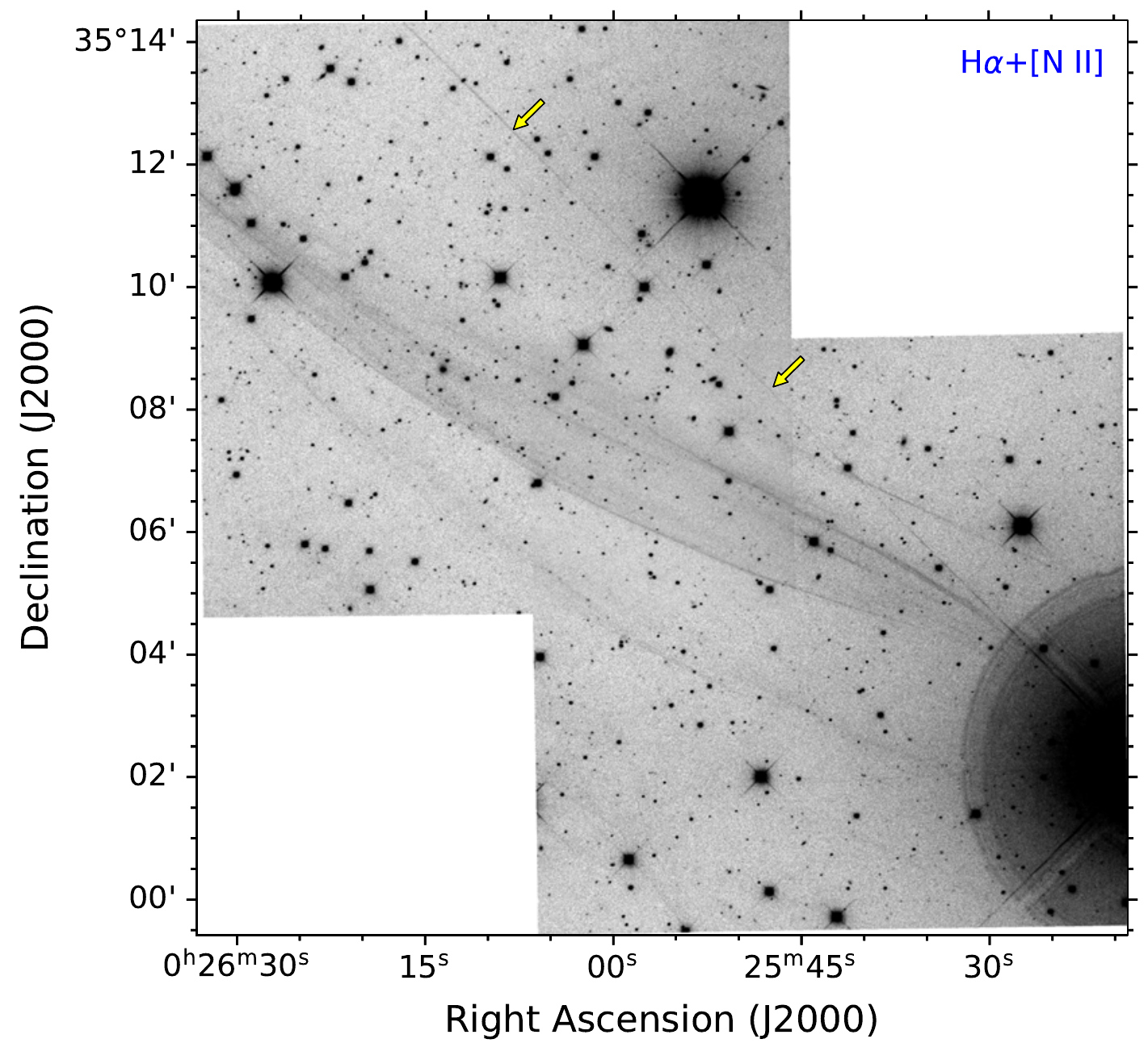}
	\caption{Two-frame mosaic in \Hafilt\, of the brighter segments in region LF. Several thin filaments
		arranged along NE-SW direction, suggesting the presence of expanding emission shells, are mixed
		with fainter diffuse light, mostly in the downstream direction of the associated shock fronts.
		It should be noticed that the perfectly straight feature, running diagonally across the
		upper-right	part of the image and marked with two yellow arrows, is neither an artificial
		satellite trail nor real thin filament. It is an optical artefact, a diffraction spike
		produced by copious light from the bright star HD 2093 (V~=~ 7.36),
		just outside the lower right edge of the image.  }
	\label{fig:LF_129cm_Ha}
\end{minipage}
}
\end{figure*}

\section{Results} \label{sec:results}
\subsection{Direct Imaging}
\subsubsection{Wide-field coverage}
 The deep (total exposure time 22.5\,h) \Hafilt\,continuum-subtracted 9-frame mosaic of the area
where candidate SNR \snr was discovered in X-rays, is shown in
Fig.~\ref{fig:Ha_mosaic_30cm}. The image center is at R.A. = $00^{\rm h}23^{\rm m}14^{\rm s}$,
Dec. = $36^{\rm d}25^{\rm m}45^{\rm s}$ and has an angular length of $4\fdg25$ on each side.
Overlaid are X-ray contours from the \textit{SRG} imaging observations \citep{Churazov_2021}.
The prominent set of two optical filaments (marked as UF -- upper filament -- and LF -- lower filament)
at the southeastern part of the X-ray image, running almost in parallel directions,
are the major features seen in this wide-field image and there is very good positional match
with the brightest contours of the X-ray map.

A few much fainter filaments are seen scattered, mostly at the eastern part of the SNR
(near the left edge of the image, at Declination ~$36\fdg5$). Two more, very faint,
straight and parallel to each other, filaments are located just north-east of the presumed
center of the remnant [at (RA, Dec) $\simeq$ ($00^{\text{h}}25^{\text{m}}$, $37\degr10\arcmin$)].
Another wider diffuse stripe is extending radially outwards the north-western corner of the image,
which is also barely visible in the DSS red images and in the All-Sky \ha
map\footnote{\url{https://lambda.gsfc.nasa.gov/product/foreground/fg_diffuse.cfm}}
with $6\arcmin$ resolution \citep{Finkbeiner_2003}. This latter feature seems to be unrelated to the SNR
emission. We did not investigate these fainter structures further, but will be included
in our follow-up observations.

Fig.~\ref{fig:Flm_30cm} presents a close-up (but still in low angular resolution) of the filaments,
in \Hafilt, \SIIfilt and \OIIIfilt. UF is brighter in \Hafilt and \SIIfilt and marginally visible
in \OIIIfilt.
It appears slightly curved at its brighter portion, with the convex part facing towards the center
of the SNR, and extends (in the red emission lines) for $\simeq \,38\arcmin$.
On the other hand, in the \OIIIfilt light, UF appears fainter but seems it can be traced
to a longer extend, getting a bit brighter towards its southwestern projected end, to the west
of star HD 2154 (marked in bottom panel of Fig.~\ref{fig:Flm_30cm}).

The lower filament (LF) is more elongated compared to UF, with an estimated angular extend at least
$70\arcmin$ (measured in the \SIIfilt image), and despite its high angular length,
it appears to be very straight, indicating it is part of a large-radius expanding shell and minimal
interaction with any large-scale denser gaseous bodies (diffuse or molecular clouds).
If we trace visually the lower filament, starting from its brighter
projected end next to HD 2093 (see top panel in Fig.~\ref{fig:Flm_30cm}), its brightness remains
more-or-less unchanged in all three emission lines, for about one third of its angular length,
where a bifurcation occurs and \OIIIfilt emission is dimmed heavily. The northern branch,
much fainter and sharper than the lower one, follows a remarkably straight path and is visible
in both \Hafilt and \SIIfilt light. The lower branch appears brighter, a little wider
and its continuity is interrupted for a short distance before its brightness weakens completely.
South of the brighter part of LF and close to it (angular separation $\lesssim 6 \arcmin$) a few short
and fainter filaments are seen, running along in parallel to the major feature.
Another network of curved filaments is visible -- mostly in \Hafilt\ -- to the west of HD 2093
(top panel in Fig.~\ref{fig:Flm_30cm}).

We summarise in Table~\ref{tab:other_flms} the locations of fainter filamentary
structures appearing in our wide-field imagery, which were not investigated any further in this
preliminary report. The celestial coordinates of each filament's estimated center along with
their approximate angular length (in arcminutes) are listed there. In the last two columns of
Table~\ref{tab:other_flms}, the corresponding Figure number and the filter used in obtaining
the associated image are given as well.

Figure~\ref{fig:Flm_30cm_color} shows in colour representation the spatial distribution of the line
emissions in \Hafilt (red), \SIIfilt (green) and \OIIIfilt (in blue hues). The shock fronts
associated with the two filaments in this picture, are moving towards the lower left direction.
In the UF, both \Hafilt and \SIIfilt contribute in brightness, with an apparent spatial precedence
of the \ha emission. No appreciable light is seen in \OIIIfilt there. However, evidence for
\OIIIfilt emission is provided both in the bottom frame of Fig.~\ref{fig:Flm_30cm} and, more convincing,
in the spectrum taken in region UF (see Sec.~\ref{sec:spectra}). Also, in the
bright section of LF, the \OIIIfilt emission lags behind the \ha emitting front.

\subsubsection{Higher angular resolution imaging}
The two major filaments detected in our deep wide-field images can now be seen in greater detail in
higher resolution \Hafilt mosaics, in Figs.~\ref{fig:UF_129cm_Ha} and \ref{fig:LF_129cm_Ha}.
A multitude of thin emission filaments exist, arranged almost in parallel
directions with no apparent major distortion through interactions with the surrounding medium
or signs of recent strong large-scale instabilities.
Fig.~\ref{fig:UF_129cm_SII} shows the \SIIfilt emission at the region covered by the middle
frame in the \Hafilt mosaic in Fig.~\ref{fig:UF_129cm_Ha}. Only the brighter filaments present
in the \ha image are seen in \SIIfilt, where the intensity scale-length in the
emitting shell behind the shock front seems to be greater than that in \ha.
This is clearly depicted in Fig.~\ref{fig:UF_129cm_HaSII}, where the \ha (red)
emission brightness falls-off more quickly behind the shock than \SIIfilt does.

At the brighter part of emission in the LF region a similar picture of several thin \Hafilt
filaments in close spacing arrangements is shown in Fig.~\ref{fig:LF_129cm_Ha}. Near the southern
frame's lower edge, a short segment of the smaller companion filament is also visible
(see Figs. \ref{fig:Flm_30cm} and \ref{fig:Flm_30cm_color}).
\begin{figure}
	\centering\includegraphics[width=0.97\linewidth]{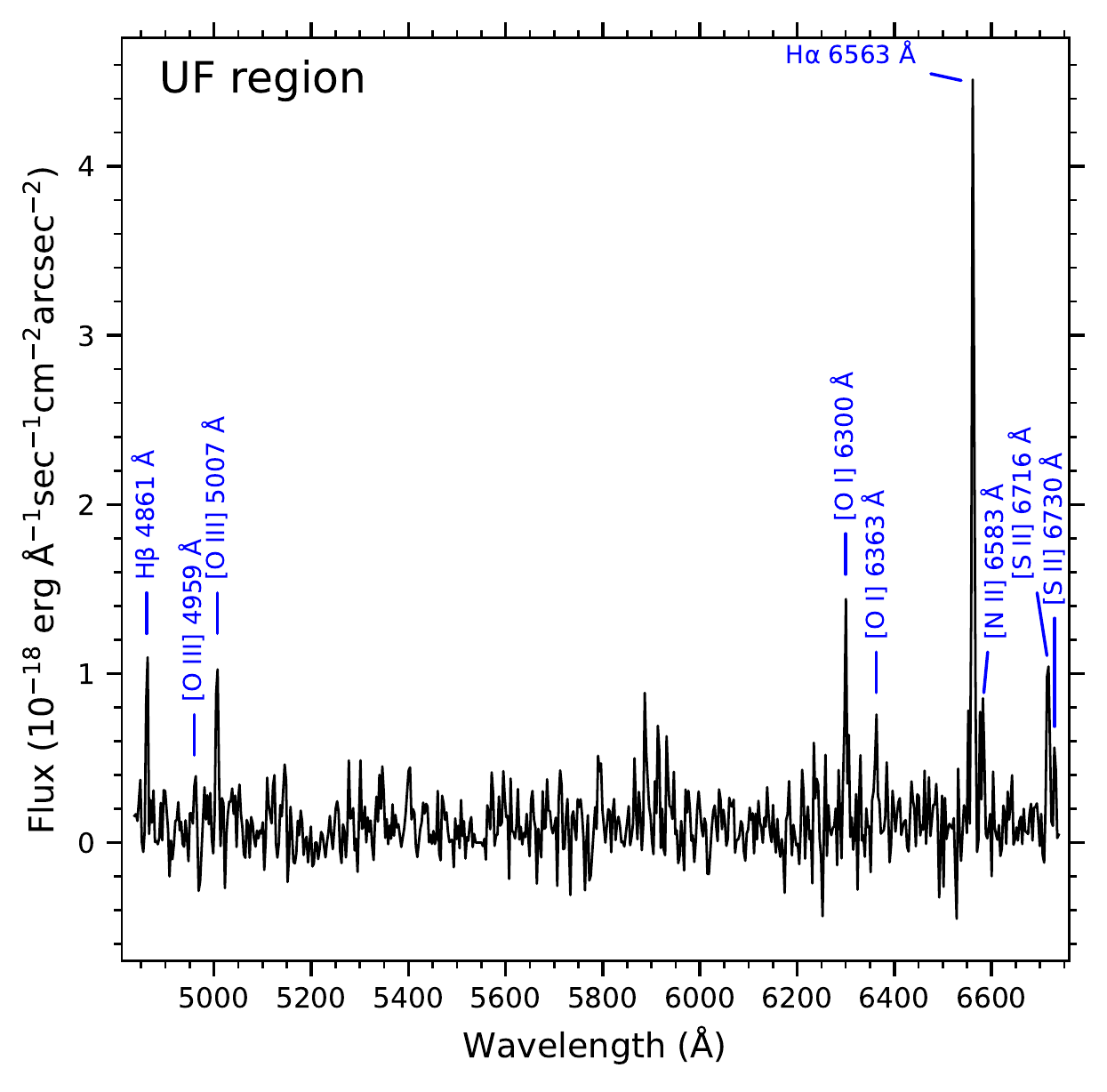}
	\caption{Emission line spectrum across filaments in region UF.}
	\label{fig:129cm_spectra}
\end{figure}
\begin{figure}
	\centering
	\includegraphics[width=0.98\columnwidth]{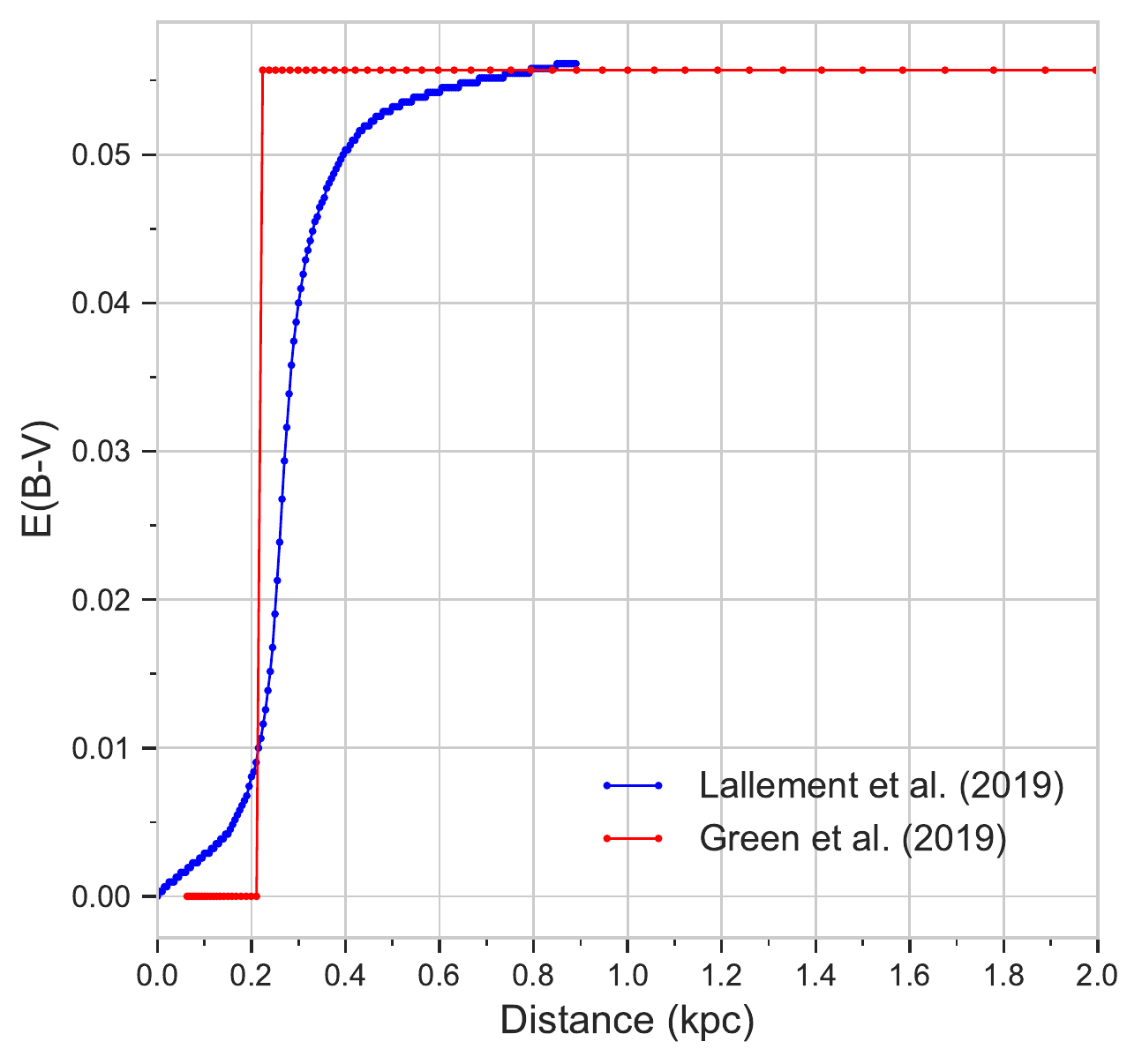}
	\caption{The colour excess $E(B-V)$ as a function of distance towards the region UF in \snr,
		according to the 3D dust maps of \citet{Lallement_2019} and \citet{Green_2019}.}
	\label{fig:extinction}
\end{figure}

\begin{table*}
	\centering
	\begin{minipage}[t]{0.58\textwidth}
	\begin{threeparttable}[b]
		\caption{Normalised emission line fluxes from the extracted spectrum in the UF filament.}
		\begin{tabular*}{\textwidth}{l@{\extracolsep{\fill}}cccc}
			\hline
			${\rm \uplambda}$ (\AA),\ ID & Extinction function, $\text{f(}\uplambda)$\tnote{a} &
			F($\uplambda$)\tnote{b} &  I($\uplambda$)\tnote{c} &  S/N\tnote{d}\\
			\hline
			4861 ${\rm H\upbeta}$   &  0.0000 & 100.0 & 100.0 &  8.1 \\ 
			4959 $[\ion{O}{iii}]$ & -0.0297 &  51.3 &  51.1 &  3.5 \\
			5007 $[\ion{O}{iii}]$ & -0.0446 & 123.6 & 122.6 &  7.7 \\
			6300 $[\ion{O}{i}]$   & -0.2835 & 151.3 & 143.6 &  8.1 \\
			6364 $[\ion{O}{i}]$   & -0.2924 &  55.2 &  52.3 &  6.1 \\
			6563 ${\rm H\upalpha}$  & -0.3224 & 382.1 & 360.0 & 39.2 \\
			6583 $[\ion{N}{ii}]$  & -0.3256 &  75.9 &  71.4 & 12.1 \\
			6716 $[\ion{S}{ii}]$  & -0.3458 & 138.5 & 129.9 &  8.7 \\
			6731 $[\ion{S}{ii}]$  & -0.3479 &  75.5 &  70.8 &  5.2 \\
			\hline
			Observed ${\rm F(H\upbeta})$\tnote{e} & & $7.75\,\pm\,0.9$ & & \\
			${\rm c(H\upbeta)}$\tnote{f} & & $0.080\,\pm\,0.17$ & & \\
			${ E(B-V)}$\tnote{g} & & $0.054\,\pm\,0.12$ & & \\
			\hline
		\end{tabular*}
	\label{tab:spectra_fluxes}
		\begin{tablenotes}
			\item[a] Derredening function according to \citet{Fitzpatrick_2019} extinction law and extinction
			ratio R(V)~=~ 3.1. See Appendix \ref{appendix:int_ext_correction} for details.
			\item[b] Observed line flux normalised to ${\rm H\upbeta = 100}$.
			\item[c] Intrinsic flux ratio, i.e. flux relative to ${\rm H\upbeta = 100}$ corrected for
			interstellar extinction.
			\item[d] Signal to noise ratio (S/N) for the observed flux ratio. Error contribution from
			the ${\rm H\upbeta}$ flux is not included in the noise term.
			\item[e] In units of \fluxu.
			\item[f] Logarithmic ${\rm H\upbeta}$ extinction coefficient: 
			${\rm c(H\upbeta) = log[(H\upbeta)_{intrinsic}	/ (H\upbeta)_{observed}]}$.
			\item[g] $B-V$ colour excess.
		\end{tablenotes}
	\end{threeparttable}
\end{minipage}
\hfill
\begin{minipage}[t]{0.38\textwidth}
	\caption{Major dereddened flux ratios of the UF spectrum.}
		\begin{tabular*}{\textwidth}{l@{\extracolsep{\fill}}c}
			\hline
			Line ratio & Value \\
			\hline
			[\ion{S}{ii}]\,$\uplambda\uplambda$\,6716, 6731\,/ \ha & $0.56\,\pm \,0.06$ \\
			$[\ion{N}{ii}]$\,$\uplambda$ 6583\,/ ${\rm H\upalpha}$ & $0.20\, \pm \,0.02$ \\
			$[\ion{O}{iii}]$\,$\uplambda$ 5007\,/ ${\rm H\upbeta}$ & $1.23\, \pm \,0.22$  \\
			$[\ion{O}{i}]$\,$\uplambda$ 6300\,/ ${\rm H\upalpha}$ & $0.40\, \pm \,0.05$  \\
			${\rm I(6716) / I(6731)}$ & $1.83\, \pm \,0.41$\\
			$[\ion{N}{ii}]$\,$\uplambda$\,6583\,/ [{\ion{S}{ii}}]\,$\uplambda\uplambda$\,6716, 6731\ & $0.36\, \pm \,0.05$\\
			\hline
		\end{tabular*}
		\label{tab:line_ratios}
\end{minipage}
\end{table*}

\subsection{Optical spectroscopy} \label{sec:spectra}

A low resolution, long-slit spectrum was acquired at a bright filament position in region UF (see
Table~\ref{tab:spectra_info}) with the Skinakas spectrograph, mount on the 1.29\,m RC telescope.
Individual exposures on the target were aligned spatially (via spectra of stars in the slit)
and along the dispersion direction (using sky emission lines), corrected for atmospheric extinction
and the 2D images were averaged. Extraction aperture was selected to sample relatively brighter
region across the filaments and background was estimated from emission free regions,
at the closest possible distance from the extraction aperture. The exact location of the
slit aperture used for object and background extraction is shown overlaid in
Fig.~\ref{fig:UF_129cm_HaSII}. Line fluxes were measured fitting Gaussians over
a linear local background estimate.

The resulting flux-calibrated spectrum is presented  in Fig. \ref{fig:129cm_spectra}.
Correction of the line emission fluxes for attenuation effects due to interstellar extinction
was performed with the Balmer decrement method.
We used the \citet{Fitzpatrick_2019} extinction law corresponding to the mean galactic ratio of
total-to-selective extinction R(V) = 3.1, and details on the derivation of the extinction function
are given in Appendix \ref{appendix:int_ext_correction}.

The usual practice in performing flux dereddening is to select, as a first step, a value
for the intrinsic Balmer decrement $\text{I(H}\upalpha)/\text{I(H}\upbeta)$, appropriate for the type
of gaseous nebula studied. In the SNR case, the adopted
values are either $\text{I(H}\upalpha)/\text{I(H}\upbeta)$\,=\,2.86  -- the theoretical
Case B recombination (nebula optically thick to the Lyman series photons) -- seen mostly in
\ion{H}{ii} regions,
or $\text{I(H}\upalpha)/\text{I(H}\upbeta)$\,=\,3.0 when conditions favour the likelihood of significant
collisional excitation of hydrogen in the post-shock region \citep[e.g.][]{Raymond_2020, Fesen_2021}. Indeed,
models with self-consistent treatment of the precursor ionization for a wide range
of shock velocities (v$_{s} = $ 10--1500~km\,$s^{-1}$) indicate that the intensity
ratio $\text{I(H}\upalpha)/\text{I(H}\upbeta)$ exceeds always the Case B value 2.86 and varies
between $\sim$ 3.0--5.0. In lower-velocity shocks the \ha/\hb flux ratio is strongly enhanced by
collisional excitation in the postshock gas \citep{Raymond_1979, Chevalier_1980, Cox_1985, Hartigan_1987,
Sutherland_2017, Dopita_2017}.
The derived value of the extinction $A(V)$ and colour excess $E(B-V)$ are then compared
to estimates from independent sources.
\begin{figure}
	\centering\includegraphics[width=0.99\linewidth]{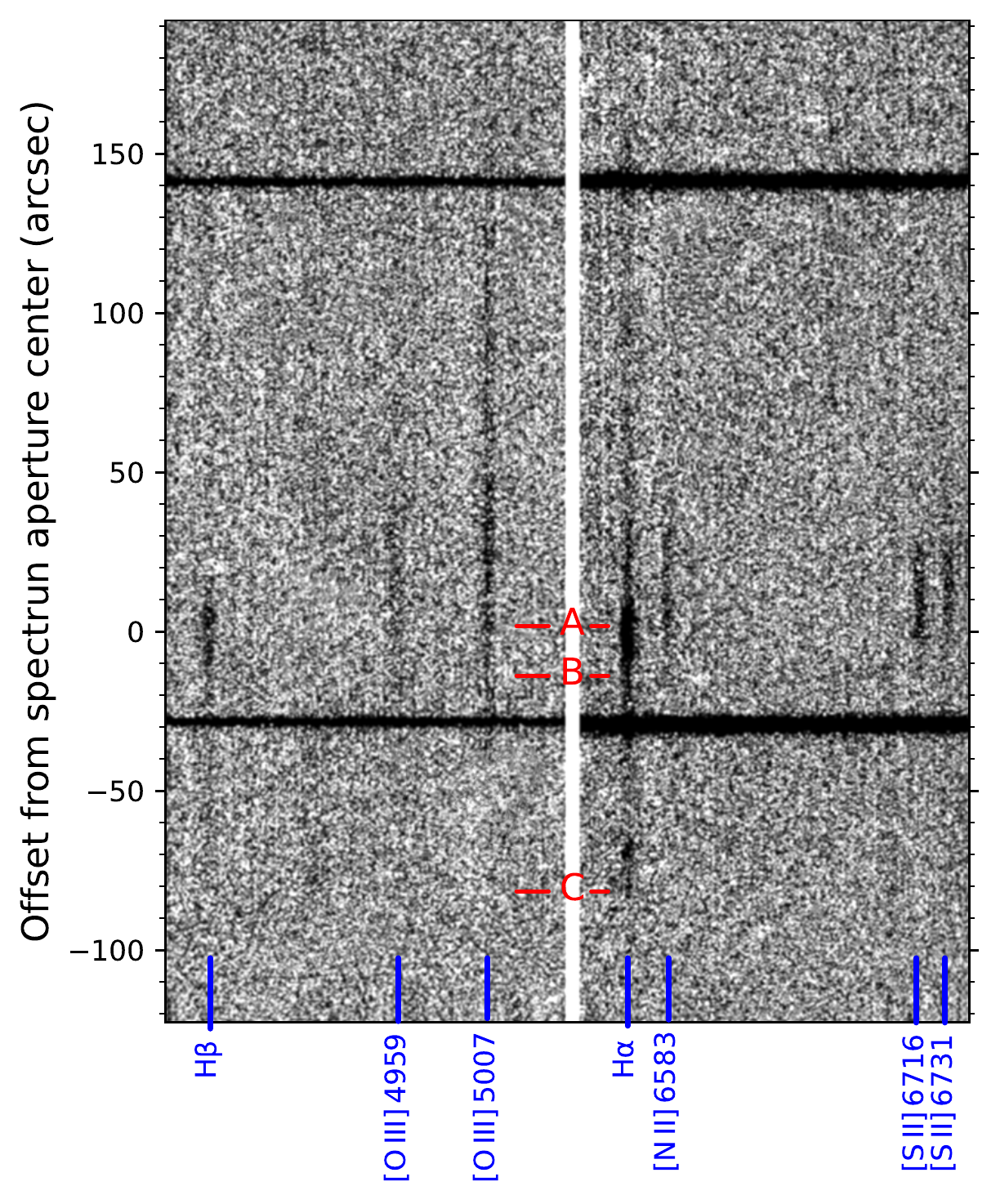}
	\caption{Part of the long slit spectral image in region UF, showing the spatial correlation of
		major emission lines identified with labels at the bottom.
		The [\ion{O}{i}] $\uplambda\uplambda 6300, 6364$ lines are not included
		for reasons explained in the text.
		The vertical axis scale corresponds to the angular offset from the extraction aperture	center
		(value 0) across the slit aligned along the N-S direction. Positive numbers correlate
		with northern displacements (post-shock region). The white vertical stripe separates
		the bluer (left) from the redder wavelengths and the two horizontal black bands running
		across the image width are spectra of stars aligned in the slit.
		Letters and short horizontal line segments in red indicate the location of filaments discussed
		in the text.}
	\label{fig:spectrum_image}
\end{figure}

In this study, we decided to take a different approach and adopt a value for the Balmer decrement 
$\text{I(H}\upalpha)/\text{I(H}\upbeta)$ such that the resulting colour excess $E(B-V)$ agrees with the
extinctions predicted by the 3D dust maps of
\citet{Lallement_2019}\footnote{\url{https://astro.acri-st.fr/gaia_dev/}} and \citet{Green_2019}
\footnote{\url{http://argonaut.skymaps.info/query}} in the direction of \snr. Both dust maps show a
gradual increase of $E(B-V)$ to 0.055 at a distance $\sim400$\,pc, and remains flat to larger distances
(see Fig. \ref{fig:extinction}). This requirement leads to intrinsic Balmer decrement
$\text{I(H}\upalpha)/\text{I(H}\upbeta)\,=\,3.6$, which corresponds to shock velocity range 
75~$\lesssim$~ (v$_{s}$\,/\,km\,s$^{-1}$) $\lesssim$ 120, as predicted by relevant models
\citep[see][fig.~18]{Sutherland_2017}. It should be emphasized that the emission line ratios,
useful in diagnostic criteria, are insensitive to the reddening parameters used, because by construction,
they correspond to wavelengths close to each other, and thus line fluxes are almost equally affected
by interstellar extinction.

Dereddened fluxes were calculated with the aid of Equations \ref{eq9} and \ref{eq8}, derived in
Appendix~\ref{appendix:int_ext_correction}, and the colour excess applying Eq.~\ref{eq11}.
Listed in Table~\ref{tab:spectra_fluxes} are the measured $\text{F}(\uplambda)$ and extinction
corrected $\text{I}(\uplambda)$ fluxes, normalised to $\text{F(H}\upbeta)=100$\ and
 $\text{I(H}\upbeta)=100$, respectively, as well as the logarithmic H$\upbeta$\, extinction
 coefficient c(H$\upbeta$) and colour excess $E(B-V)$.
The S/N values given in the last column of Table~\ref{tab:spectra_fluxes} do not include
calibration errors, estimated to be less than 15 per cent, neither the contribution of
the normalising H$\upbeta$\, flux error.
It should be also noticed the relatively high uncertainty in the interstellar extinction
related values of c(H$\upbeta$) and $E(B-V)$, reflecting the high relative error in the
\hb flux (8 per cent).

The spectrum of a filament in region UF presents clear evidence that the associated emission results
from shock-heated gas, commonly seen in SNRs. The widely accepted criterion supporting this is the
observed value of [\ion{S}{ii}]\,/\,H$\upalpha$\,=\,0.56\,$\pm$\,0.06, exceeding the threshold 0.4,
which separates SNRs from \ion{H}{ii} regions
\citep{Dodorico_1978, Blair_1981, Dopita_1984, Fesen_1985}.
Another sensitive shock emission test is provided through the [\ion{O}{i}]\,/\,H$\upbeta$ ratio,
found to be 0.40 at a significance level of $8\,\upsigma$. This line ratio does not exceed 0.1 in
\ion{H}{ii} regions while higher values are associated with emission from shock heated interstellar
nebulosities, like the shocks in SNRs \citep[e.g.][]{Fesen_1985, Kopsacheili_2020}.

The electron density-sensitive [\ion{S}{ii}] ratio $\uplambda\uplambda$ 6716\,/\,6731 was found
to be above the low-density limit, [0.436 $\le$\,6716\,/\,6731\,$\le$\,1.434
\citep{Proxauf_2014, Osterbrock_2006}], but with a high uncertainty,
${\text{ I(6716) / I(6731)}}$ = $1.83\, \pm \,0.41$. Therefore, we can not estimate the electron
density (in the zone where S$^{+}$ recombines) with confidence but only state that
n$_{e}\,\lesssim100$\,cm$^{-3}$. 

Another remarkable property revealed from the spectrum sampled in the UF filament is its
low brightness. The measured flux F(H$\upalpha$)~=~(2.96\,$\pm$\,0.08)\,
$\times 10^{-17}$\,erg\,\AA$^{-1}$\,s$^{-1}$\,cm$^{-2}$\,arcsec$^{-2}$ is 1--3 orders of magnitude
fainter than most of the Galactic SNRs observed in the optical \citep[e.g.][]{Alikakos_2012, Boumis_2005,
Boumis_2008, Boumis_2009, Mavromatakis_2002, Mavromatakis_2005, Mavromatakis_2009, How_2018}.

\subsection{Post-shock emission morphology}

Figure~\ref{fig:spectrum_image} presents two sections of the negative, background-subtracted 2D spectral
image in the region UF filaments. The spatial variation of emitted fluxes in several emission 
lines -- marked at the bottom of the picture -- is shown with respect to angular distance
from the aperture center (at value 0) used for the spectrum extraction, along the slit, aligned in the
N-S direction. Positive numbers on the left axis scale correspond to northern displacements
(down-stream direction relative to the shock fronts).
Marked also with letters and short horizontal red line segments are the locations of filaments,
as identified from the higher resolution \Hafilt image in Fig. \ref{fig:UF_129cm_Ha}.
It should be mentioned that only light emitted from filaments A and B
is included in the aperture of the extracted spectrum in Fig.~\ref{fig:129cm_spectra}.
\begin{figure*}
	\begin{minipage}[t]{0.49\textwidth}
		\includegraphics[width=0.99\columnwidth]{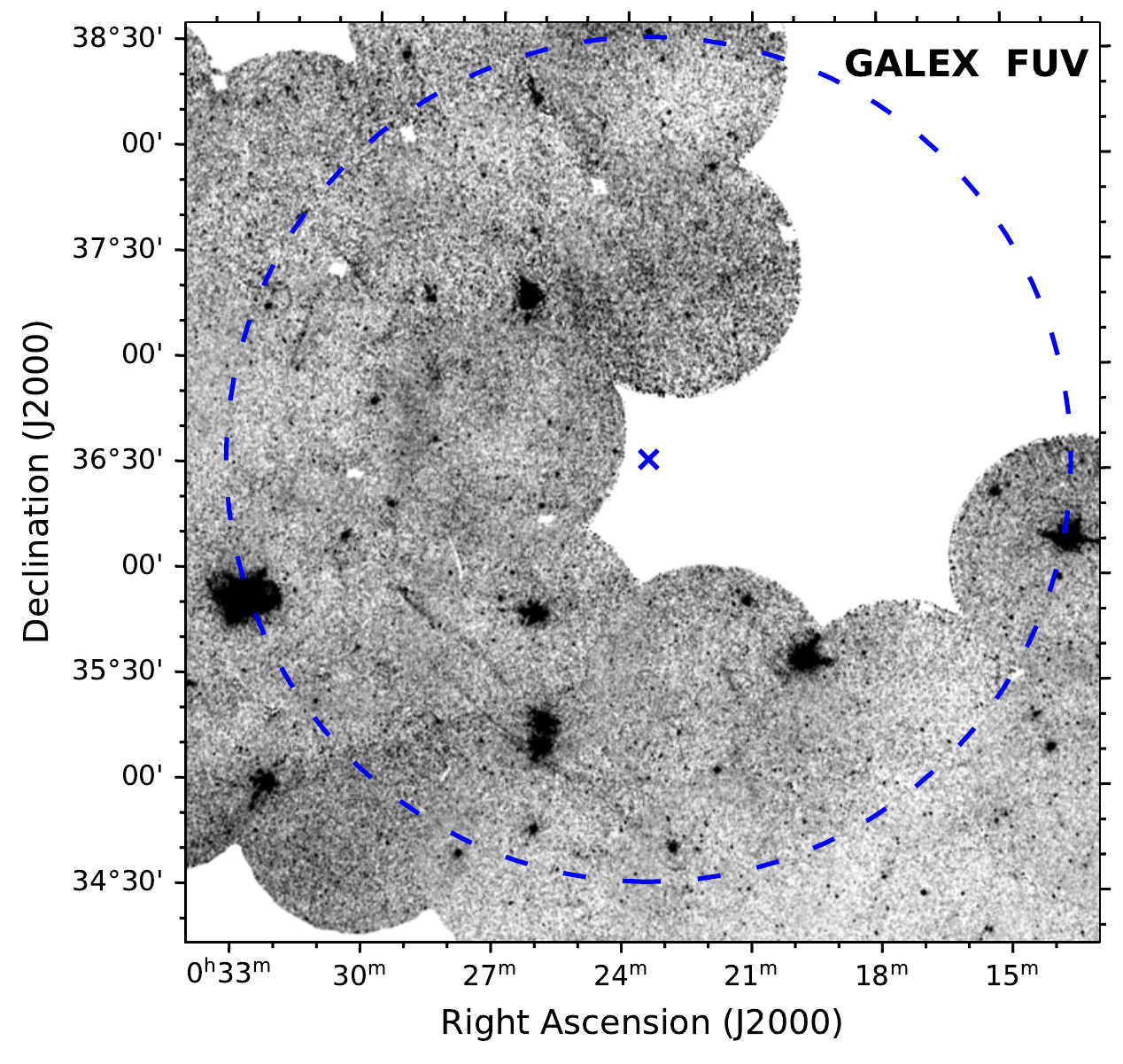}
	    \caption{GALEX-FUV image covering the area of \snr. The dashed blue circle defines the X-ray
	    extend of the supernova remnant and the $\times$ symbol marks its center.
	    The filamentary structures near the center of the south-eastern quadrant are located
	    at the same position as in our \Hafilt wide-field image in Fig.\ref{fig:Ha_mosaic_30cm}.}
	    \label{fig:GALEX_FUV}
	\end{minipage}
\hfill
{
	\begin{minipage}[t]{0.49\textwidth}
		\includegraphics[width=0.99\columnwidth]{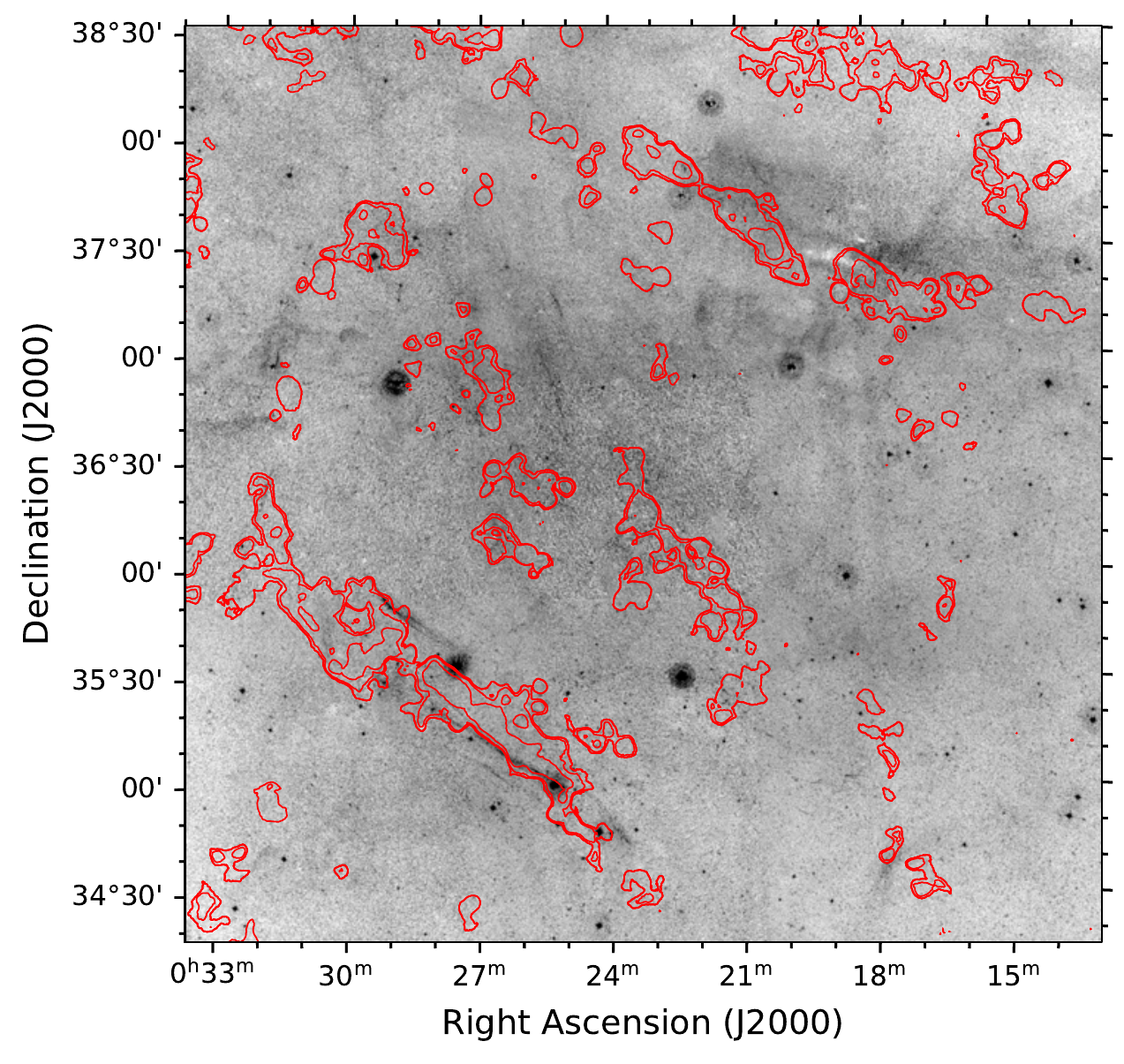}
	    \caption{Contours of radio emission in 120--168 MHz from LOFAR LoTSS DR2 images superposed on
	    our continuum-subtracted wide field \Hafilt image (Fig. \ref{fig:Ha_mosaic_30cm}).
	    The coincidence of the major optical filaments with the enhanced radio brightness at the
	    south-eastern edge is evident. Contours shown correspond to 1, 5 and 20 $\upmu$Jy\,beam$^{-1}$
	    levels of flux density in the radio data.}
	    \label{fig:LOFAR_contours}
		\end{minipage}
}
\end{figure*}

Despite the fact that [$\ion{O}{i}$]\,$\uplambda\uplambda$\,6300,6364 emission lines are equally
important in SNRs, we do not include the corresponding part in the 2D spectral image in Fig. 10.
The reason is that these lines are also quite strong in the night sky spectrum and the background
subtraction procedure does not provide a smooth result at the positions of the [$\ion{O}{i}$] lines.
Thus, in our low resolution spectroscopy, the absence of considerable velocity component
in the SNR [$\ion{O}{i}$] lines to produce sufficient Doppler shift, combined with the intrinsic
faintness of the nebular emission (even in H$\upalpha$) yield relatively \lq noisy\rq\ 
background-subtracted [$\ion{O}{i}$] lines and result in a rather confusing picture.

Starting from the most southern filament C, this appears as a diffuse and very faint feature which is
seen in the spectrum to emit basically \ha light and just a weak hint of \hb emission
at nearly the noise level.
Filaments A and B are projected close to each other, at an angular distance $\simeq 19\arcsec$.
\ha emission starts ahead of the brightest parts of the filaments, along with some \hb and weak
\OIIIfilt\ light, but no detection of emission from other low-excitation lines, such as
[$\ion{N}{ii}]$\,$\uplambda\uplambda$\,6548,6583 and [$\ion{S}{ii}$]\,$\uplambda\uplambda$\,6716,6731.
The \ha intensity increases for a short distance and then slowly decreases over a relatively
long interval ($\simeq\,100\arcsec$) before it fades completely.
The second Balmer line (\hb) follows a similar variation but becomes undetectable much earlier,
shortly after its peak. Heavier ions, like N$^{+}$\, and S$^{+}$, start emitting further downstream,
behind the location of maximum \ha emission and for a short angular distance, roughly 30$\arcsec$.

In contrast, in the line \OIIIfilt~$\uplambda$\,5007, emission can be traced for much longer distance,
seen almost as far as the \ha light, but in low intensity. The detected \OIIIfilt\ emission
in the post-shock region suggests a lower limit on the shock velocity v$_{s}$ of about
70~km\,s$^{-1}$, while its low brightness implies that this value is not expected to be much
higher than 90 km\,s$^{-1}$. These limits were found from the diagram in fig. 6 of
\cite{Dopita_2017}, using the measured line ratio
log([$\ion{O}{iii}$]\,$\uplambda$5007\,/\,H$\upbeta$) = 0.09\,$\pm$\,0.08
(see Table~\ref{tab:line_ratios}), and come to agreement with the implied shock velocity range
through our earlier choice for the intrinsic Balmer decrement $\text{I(H}\upalpha)/\text{I(H}\upbeta)$,
in the extinction correction procedure of the spectral line fluxes. 

One additional advantage of using the line ratio [$\ion{O}{iii}$]\,$\uplambda$5007\,/\,H$\upbeta$
to constrain the shock velocity range is the very weak dependency of its value on density
\citep{Dopita_2017}. However, the lower limit on the shock velocity implied by the detection of
[$\ion{O}{iii}$]\,$\uplambda$5007\, emission, is based on the assumption that O$^{++}$ ions
are not present in the preshock gas, or else [$\ion{O}{iii}$] emission lines can be produced
in much slower shocks \citep{Raymond_1979}. This latter situation cannot be excluded from our spectrum,
since the preshock area is projected on the postshock region of another, much weaker shock-front,
leading ahead of the brighter filament in this region.

A similar structure behind relatively slow shocks (v$_{s}\lesssim$\,150~km\,s$^{-1}$) was found
in optical spectroscopic investigations of other high galactic latitude SNRs, such as G\,70.0$-$21.5
\citep{Raymond_2020}, G\,107.0$+$9.0 \citep{Fesen_2020} and in the Hoinga or G\,249.5$+$24.5 SNR
\citep{Becker_2021, Fesen_2021}. The coexistence of radiative and Balmer dominated shock fronts in these
supernova remnants indicate variable velocities of expansion in filamentary structures
at different locations of the remnants.

\section{ Observations at other wavelengths} \label{sec:other_wavebands}
\subsection{X-rays} \label{sec:x-rays}
\snr has been discovered in X-rays, through the \textit{SRG}/eROSITA all-sky survey
\citep{Churazov_2021}, as a faint, very extended circular feature with a total 0.3--2 keV
flux of about 3\,$\times$\,10$^{-11}$~erg s$^{-1}$cm$^{-2}$ distributed over a projected area
$\sim$\,11.8~deg$^{2}$. The soft X-ray image depicts the circular source with a slightly brightened
rim, especially across the south-eastern edge and two compact knots (seen in the smoothed version
of the X-ray image), one near the western ridge and the other close to the center of the object.
Comparing the location of the two major optical filaments in Fig.~\ref{fig:Ha_mosaic_30cm} with
the overlaid X-ray brightness contours, there seems to be a clear correlation of the optical
features with one of the stronger X-ray emission parts of the remnant at south-west. A second area
of enhanced X-ray brightness appears at the northern part of the remnant, where no optical
counterpart was found thus far.

\subsection{Ultraviolet data} \label{sec:UV}
Our search for ultraviolet counterpart to the optical filaments has been performed with the publicly
available imagery of the \textit{Galaxy Evolution Explorer} (GALEX) space telescope
All-Sky Imaging survey (AIS), described in detail in \cite{Morrissey_2007}. 
The GALEX archival database contains far-UV (FUV, $\mathrm{\uplambda_{eff}\sim}$1528~\AA, 1344--1786~\AA)
and near-UV (NUV, $\mathrm{\uplambda_{eff}\sim}$2310~\AA, 1771--2831~\AA) direct imaging and
grism spectra.

Figure \ref{fig:GALEX_FUV} presents the GALEX FUV intensity mapping of the SNR \snr area,
enclosed in a dashed circle of radius 2$^{\circ}$ determined from the X-ray discovery image.
Evidently, the two parallel filaments present in the middle of the south-east quadrant in the
FUV image, are registered at exactly the same location as in our optical image and have similar lengths.
These filaments are also visible in the GALEX NUV image, although they appear quite fainter.
The NUV image is not shown here because the numerous artefacts in it produce a rather confusing
picture of the field. One more, much shorter but quite sharp filament can be seen
near the eastern edge of the FUV image (Declination $37\degr 00\arcmin$--$37\degr 20\arcmin$),
which is not clearly seen in our wide field \Hafilt image, a point that will be re-examined during
our follow-up observations.

\subsection{Observations in Radio frequencies} \label{sec:radio_obs}
We have also used archival data from the large scale survey GALFA-$\ion{H}{i}$ DR2 of Galactic neutral
hydrogen in the 21 cm radio line \citep{Peek_2018}, to search for possible radio emission in the
supernova remnant area, partially included in declination within the field covered by the survey.
The available data cover a velocity range from $-$188.4 to $+$188.4 km s$^{-1}$ with respect to the
Local Standard of Rest, in slices with spacing 0.184 km s$^{-1}$. Several features are present
in the area but no association with the visible filaments or the overall X-ray contour
morphology could be established.

However, \citet{Churazov_2022} reported recently detection of extremely faint radio emission
towards \snr from the LOFAR Two-metre Sky Survey (LoTSS) DR2 images \citep{Shimwell_2022}.
These data include 120--168~MHz images, mapping 27 per cent of the northern sky in low frequency
at 6$\arcsec$ resolution, offering full bandwidth Stokes \textit{I} total intensity maps,
linear polarization image cubes and circular polarization continuum images.

In order to examine the spatial correlation between the optical filaments with the radio emission,
we constructed a mosaic from archival LOFAR images of the \snr
projected area of size $5^{\circ} \times 5^{\circ}$, partially removed point-like radio sources and
created a network of contours at flux density levels 1, 5 and 20 $\upmu$Jy\,beam$^{-1}$.
Fig.~\ref{fig:LOFAR_contours} shows these radio contours overlaid on our continuum subtracted \Hafilt
wide field image (in Fig.~\ref{fig:Ha_mosaic_30cm}). The major parts of the radio emission structure
consist of two elongated linear features, almost parallel to each other and running along the NE to SW
direction, located at the diagonals of the SE and NW quadrants of the image.
These linear structures' northern ends are connected by a faint circular arc, completing a figure
reminiscent of a horse-shoe. The eastern branch of this feature is the brightest and coincides
with the location of the two major optical filaments, as seen in Fig.~\ref{fig:LOFAR_contours}.

The north-western segment may indicate the location for another possible optical feature.
Close inspection of our \Hafilt image reveals the presence of some diffuse and extended optical
emission, displaced towards the NW relative to the radio contours, making the association between
radio and optical emission questionable. It should be also noticed that in this area material of
elevated opacity is present -- indicated by the white shade in the negative representation
of the \Hafilt image -- which, if located in the foreground of the remnant, makes any possible
optically enhanced emission in this region hardly visible. 

\subsection{Infrared and sub-millimetre wavelengths}
\cite{Churazov_2021} presented the \textit{IRAS} $100 \micron$ image of the \snr area and compared
the infrared emission map to the X-ray structures, noticing the existence of a partial anticorrelation
between the two. They raised the possibility that the dust distribution may have been
influenced by \snr, in which case the dust and the remnant are located roughly at the same distance
from us, about 300 pc. Based on possible future confirmation of this neighbouring, the authors
propose an alternative scenario for \snr being a nearby, core-collapse SNR. Our preliminary data do
not suffice to confirm or disprove this possible scenario.

We conducted an extensive search for counterparts at other infrared and sub-millimeter wavebands, but
did not find any signs of morphological similarity with either the optical or X-ray structure of \snr.
The data used for the search stem from surveys such as the \textit{WISE All-Sky}
(all four bands, at 3.4, 4.6, 12 and 22 $\micron$), the
\textit{AKARI Far-infrared All-Sky Survey} (the WideS band with centre wavelength 90 $\micron$),
\textit{Plank Legacy Archive} (at frequencies 143, 217, 353, 545 and 857 GHz and the CO[$1\rightarrow0$],
CO[$2\rightarrow1$], CO[$3\rightarrow2$] emission maps derived from the 100, 217 and 353 GHz channels
of the Planck High Frequency Instrument).

\section{Discussion} \label{sec:discussion}

As already noted in Section~\ref{sec:intro}, \snr belongs to the few SNRs at high Galactic latitudes
(|b|\,>\,15$^\circ$) that have been observed and detected in the optical, along with other wavelengths.
These are G\,70.0$-$21.5 \citep{Boumis_2002, Fesen_2015, Raymond_2020},
G\,249.7$+$24.7 \citep[the so-called Hoinga SNR,][]{Becker_2021, Fesen_2021},
G\,275.5$+$18.4 \citep[Antlia SNR,][]{McCullough_2002, Shinn_2007, Fesen_2021}
and  G\,354.0$-$33.5 \citep{Testori_2008, Fesen_2021}.
Given their large size, faint emissivity and estimates for low shock velocities
($\sim$ 70--100~km\,s$^{-1}$), these SNRs seem to go through their mature phases of their lives.

The current work comes to present evidence, for the first time, for the existence of optical emission
from \snr. Through our imaging and spectroscopic observations,
two filamentary structures are revealed (UF and LF) in a partial shell formation. Several fine 
filaments in the UF region denote optical emission originating from shock-heated gas, usually
associated with the shock fronts of the remnants. At the same time, the compelling sequence of
emission lines can be easily traced, depicting the post-shock (emission) morphology of the source
(Figs \ref{fig:UF_129cm_HaSII} and \ref{fig:spectrum_image}). 
From the available spectrum, we see that the one out of the two strong optical filaments examined
presents a low shock velocity ($\sim$70--100 km s$^{-1}$). The latter, in conjunction with the
referred X-ray/radio angular size, seems to also place \snr in the class of older SNRs. 

\snr was primarily identified as a potential SNR on the basis of its soft X-ray properties
\citep{Churazov_2021}. Sources that present low-temperature ($<$\,2~keV), emission line spectrum from
hot, ionized gas are regarded as thermal X-ray emitting SNRs \citep[e.g.][]{Garofali_2017,
Leonidaki_2010}. Optical evidence for shock-excited gas arises from the \SIIfilt / \ha emission line
ratio ($>$0.4) in one of its two distinct filaments. Further
confirmation regarding the nature of this filament comes from the fulfillment of 9 out of the 10
applicable diagnostic tests described in \citet{Kopsacheili_2020}
that contain the emission line ratios [\ion{O}{iii}] / \hb, [\ion{O}{i}] / \ha, \SIIfilt / \ha,
and [\ion{N}{ii}] / \ha. All the aforementioned, in combination with the spatial
correlation of the optical structures with some of the brightest X-ray, UV and radio emission features
(see Section \ref{sec:other_wavebands} and Figs. \ref{fig:Ha_mosaic_30cm} and \ref{fig:LOFAR_contours})
provide robust evidence for establishing \snr as a new Galactic SNR.

Owing to their position and low surface brightness, the group of faint and large remnants at high
Galactic latitudes have escaped detection
and an in-depth study, since most of the SNR surveys focus on the vicinity of the Galactic Plane,
where wealth of dust and gas help stars to flourish and eventually die. Especially in the case of
core-collapse SNRs, this area is the ideal nursery for their birth while their existence above
the usual regions of research (|b|\,$\leq 10^{\circ}$) is less expected, since white dwarf 
binary systems, responsible for Type Ia supernovae explosions, are more likely to be found at higher
galactic latitudes. However, three out of five high galactic latitude SNRs mentioned previously,
have been classified in the literature as core-collapse SNRs.

In an attempt to cautiously advocate in favor or against a specific SNR progenitor, various criteria
should be taken into account. The basic characteristic of Type Ia SNRs comes from their spectrum: i.e.
the existence of prominent narrow and broad emission line components of \ha emission and the
absence (or sometimes weakness) of forbidden lines. This occurs because they are thought to be
surrounded by a medium consisting mainly of neutral hydrogen, producing non-radiative shocks
\citep[e.g.][]{Chevalier_1980}: i.e. shocks for which their radiative cooling time is much longer
than their characteristic dynamical time scale (e.g. age). Given the aforementioned, they are
expected to be found within regions isolated from massive stars / OB associations that ionize
the ambient medium \citep{Maggi_2016, Franchetti_2012}. Furthermore, the well-established criterion
for identifying core-collapse SNRs (\SIIfilt/\ha $>$0.4) drops down to less than 0.05 in the case
of Type Ia SNRs \citep{Lin_2020}. Another criterion that has been used to separate core-collapse
from Type Ia SNRs is their morphology. Assuming an unperturbed surrounding, it has been demonstrated
\citep{Lopez_2011} that Type Ia SNRs present a symmetric explosion and morphology.
However, the notion that Type Ia SNRs do not shape their environment has started to be challenged
the last few years and various exceptions cannot be ruled out \citep{Zhou_Vink_2018}. For example,
based on models reproducing the  observables, it has been argued that famous cases such as the
historical Kepler and Tycho SNRs, present extensive massive outflows from their progenitor binary,
strongly modifying this way their circumstellar medium \citep[e.g.][]{Chiotellis_2012, Zhou_2016}.

On the basis of the derived spectroscopic properties of the UF optical filament, \snr presents
a typical spectrum of a core-collapse SNR, with enhanced \SIIfilt / \ha emission line ratio and various,
definite forbidden lines denoting the shock excitation of the surrounding medium.
Although it is only one spectrum we have in hand, significant information can be drawn since it
is part of the two prominent optical filaments of
the entire SNR. However, the extracted \ha flux of the optical upper filament, with absolute flux
1-3 orders of magnitude lower compared to other Galactic SNRs, supports the notion of a Type Ia
progenitor \citep{Franchetti_2012}. What could further endorse this scenario is the almost
circular/symmetric morphology seen in the X-ray/radio images.
As can be realised, the existing data do not allow us to explicitly decide over one progenitor
scenario or the other. The determination of the distance of \snr (whether it is part of the disc
or the Galactic halo, for example) is still ambiguous and which is, by the way, a
well-known caveat in the research of Galactic SNRs. A twofold scenario may also be the case for \snr as
reported in other Galactic SNRs \citep[e.g.][]{Fesen_2020}: shock-heated line emission may coexist with
Balmer-dominated filaments. 

\section{Conclusions} \label{conlcusions}
In this preliminary investigation for the existence of an optical counterpart to the candidate 
SNR \snr, we successfully detected emission from related filamentary stuctures, through deep wide-field
and higher angular resolution CCD imaging of the area, followed by supplementary
spectrophotometric observations. 

First detection was achieved after the construction of a deep, wide-field image mosaic 
($4\fdg25 \times 4\fdg25$, scale ${\sim}3\arcsec$ pixel$^{-1}$)  through a narrowband \Hafilt filter,
covering the footprint of the X-ray discovering image. Two distinct, major filaments were revealed,
located at the southeastern ridge of the X-ray structure, in a partial shell-like formation.
These features were detected in \SIIfilt and \OIIIfilt emission
as well. Images acquired in higher angular resolution, mostly in \Hafilt, present networks of sharp
filaments in both areas, running almost parallel to each other with no significant signs
of disturbance from dynamical interactions with the ambient medium or internal large-scale
instabilities (indicating a rather old remnant).

Our deep flux-calibrated and dereddened low-resolution spectrum, obtained at a single location of the
northern filament, allowed us to confirm the nature of \snr as a most likely Galactic SNR. The
spectroscopic data show faint but significant emission from lines typically observed in SNR sources,
such as \ha, \hb, \OIIIfilt ($\uplambda\uplambda$\,4959, 5007), 
[\ion{O}{i}] ($\uplambda\uplambda$\, 6300, 6364),
\nii ($\uplambda$\,6583) and \SIIfilt ($\uplambda\uplambda$\,6716, 6731).
The \SIIfilt/\ha line ratio was found greater than 0.5, indicating emission from shock-heated gas,
and 9 out of the 10 applicable emission-line ratio diagnostic tests presented in
\cite{Kopsacheili_2020}, turned out to be positive. The measured \OIIIfilt/\,\hb line ratio
suggest\textcolor{magenta}{s} relatively low shock-front velocities (roughly 70--100 km\,s$^{-1}$).

Comparison with multiwavelength data confirmed collocation between the optical
filamentary signature of the SNR \snr with emission in X-rays and low-frequency radio, detected
and reported by \cite{Churazov_2021} and \cite{Churazov_2022}, respectively.
A search in the GALEX online archive helped us to identify fine FUV-emitting filaments,
which overlap almost perfectly with the brightest portions of their optical counterparts.

We also discussed possible scenarios about the nature of the progenitor star but could not reach
a definite conclusion due to luck of supporting data.

Although we have presented strong evidence that \snr belongs to the Galactic SNR group, several of its 
properties remain undetermined. Our ongoing follow-up imaging and spectroscopic investigation will help
to uncover the physical properties of the entire remnant (e.g. plasma conditions, kinematics,
environment, distance, evolutionary stage) and shed more light on its nature.

\section*{Acknowledgements}
We would like to thank the anonymous referee for his/her valuable feedback and constructive comments
that improved the content of our research paper.

This research made use of \textsc{Montage}. It is funded by the National Science Foundation
under Grant Number ACI-1440620, and was previously funded by the National Aeronautics and
Space Administration's Earth Science Technology Office, Computation Technologies Project,
under Cooperative Agreement Number NCC5-626 between NASA and the California Institute of Technology.

This publication utilizes data from Galactic ALFA HI (GALFA HI) survey data set obtained with the
Arecibo L-band Feed Array (ALFA) on the Arecibo 305m telescope. The Arecibo Observatory is operated
by SRI International under a cooperative agreement with the National Science Foundation (AST-1100968),
and in alliance with Ana G. Méndez-Universidad Metropolitana, and the Universities Space Research
Association. The GALFA HI surveys have been funded by the NSF through grants to Columbia University,
the University of Wisconsin, and the University of California.

LOFAR data products were provided by the LOFAR Surveys Key Science project (LSKSP;
\url{https://lofar-surveys.org/}) and were derived from observations with the International LOFAR
Telescope (ILT). LOFAR \citep{vanHaarlem_2013} is the Low Frequency Array designed and constructed
by ASTRON. It has observing, data processing, and data storage facilities in several countries,
which are owned by various parties (each with their own funding sources), and which are collectively
operated by the ILT foundation under a joint scientific policy. The efforts of the LSKSP have
benefited from funding from the European Research Council, NOVA, NWO, CNRS-INSU, the SURF Co-operative,
the UK Science and Technology Funding Council and the Jülich Supercomputing Centre.

This research has made use of the SIMBAD database, operated at CDS, Strasbourg, France
\citep{Wenger_2000}. 

M.K. acknowledges funding from the European Research Council under the European Union’s Seventh
Framework Programme (FP/2007-2013)/ERC Grant Agreement No 617001. She also acknowledges support
from the European Research Council under the European Union’s Horizon 2020 research
and innovation program, under grant agreement No 771282.

\section*{Data Availability}

The optical images and spectrum taken at Skinakas Observatory and used in this research
will be available upon a reasonable request to the corresponding author.



\bibliographystyle{mnras}
\bibliography{SNR_G1166-261} 




\appendix

\section{Image subtraction procedure}
\label{appendix:image_subtraction}

The image subtraction procedure described here was used \emph{only} when the
background-subtracted image was needed for visual examination and detection of possible
faint optical emission.
The continuum (C) image was registered to the narrow-band (NB) image using the \texttt{astropy}
\citep{astropy:2013, astropy:2018} package \texttt{astroalign} \citep{Beroiz_2020},
which is achieved by finding similar 3-point asterisms (triangles) in both images and estimating
the affine transformation between them.

Next we performed a background noise estimation to set the threshold pixel value for source detection
that follows. This was based on procedures of the package \textsc{SEP} \citep{Barbary_2016}, a
stand-alone \textsc{PYTHON} adaptation of the widely used program \textsc{SEXTRACTOR}
\citep{Bertin_1996} for source detection and photometry. The method used consists of the following steps,
applied to each of the two images separately.
(1) Saturated pixels are masked and a spatially variable background is fit.
(2) A  background-subtracted image is computed along with an estimate of the background noise
for each image pixel.
(3) Source detection follows with a threshold set to 3$\sigma$, where $\sigma$ is the global
root-mean-square (RMS) of the background noise.
(4) Sources are deblended and a catalogue is prepared with source centroid positions
and fluxes (equivalent to the FLUX\_AUTO in \textsc{SEXTRACTOR}), measured over elliptical apertures
centered at each detected source and corrected for local background contribution.

The sources detected in the NB and C images (already aligned at an earlier stage) were matched
on source centroid proximity criteria with \textsc{STILTS}, the command-line sister package to
\textsc{TOPCAT} \citep{TOPCAT_STILTS}. The number of matched stars per single frame 
(FoV $100\arcmin \times 100\arcmin$) was in the range of 5500--11000, depending on the narrow-band
filter type, and their flux ratio \textit{flux}(NB) / \textit{flux}(C) was computed.
This set of flux ratios was iteratively sigma-clipped, with 3$\sigma$ as the limit;
the removed outliers were mostly the brighter of the stars.
The average value of the sigma-clipped set was used to normalise the continuum image before
subtracting it from the narrow-band. The continuum-subtracted result was always inspected visually.
In a few cases, instead of the mean value, we used either the median or the 75 percentile 
of the cleaned ratios set. 

In order to reduce the effect of over- or under- subtraction of the star light and enhance visibility
of faint emission, we applied a 3$\times$3 median filter to the whole continuum-subtracted image.
When this smoothing did not produce a satisfactory result (too many 'dark holes' left in the image), 
we replaced the pixel values within the elliptical apertures -- used previously in the flux estimation --
with the median value of pixels within a narrow elliptical ring
around the aperture and an additive Gaussian noise component with local characteristics.  

\section{Interstellar extinction correction}
\label{appendix:int_ext_correction}
Emission line ratios were corrected for interstellar reddening using the normalized extinction curve of
\citet{Fitzpatrick_2019} (F19 hereafter), produced by combining \textit{Hubble Space Telescope}/STIS
and \textit{International
Ultraviolet Explorer} spectrophotometry and \textit{Two Micron All-Sky Survey} broadband NIR photometry. 
Their extinction curves are defined through the quantities:
\begin{flalign} \label{eq1}
	k(\uplambda - 55) & \equiv \frac{E(\uplambda - 55)}{E(44 - 55)} = \frac{A(\uplambda) - A(55)}{A(44) - A(55)}\\
	                & = k'(\uplambda)\quad \text{(for shorter notation)} \nonumber
\end{flalign}
\begin{equation} \label{eq2}
	\text{and}\quad R(55) \equiv \frac{A(55)}{E(44 - 55)}
\end{equation}
where $A(\uplambda)$ is the total extinction (in magnitudes) at wavelength $\uplambda$,
$E(\uplambda_{1} - \uplambda_{2})$ denotes the colour excess (that is, the difference between the observed colour
 and the expected colour in the absence of interstellar absorption) and the numbers 44 and 55 correspond to
 monochromatic wavelengths of 4400 \AA\, and 5500 \AA, respectively, instead of the most commonly used Johnson
  B and V filters.
  
If $F(\uplambda)$ and $I(\uplambda)$ denote the observed and intrinsic flux, respectively, of a nebular emission
line at wavelength $\uplambda$, then the colour excess of the Balmer decrement can be written as:
\begin{flalign} \label{eq3}
	E({\rm H\upbeta - H\upalpha}) &= A({\rm H\upbeta}) - A({\rm H\upalpha}) \nonumber \\
 	                          &= [A({\rm H\upbeta}) - A(55)] - [A({\rm H\upalpha}) - A(55)] \nonumber \\
 	                          &= [k'({\rm H\upbeta}) - k'({\rm H\upalpha})] E(44 - 55)
\end{flalign}
or
\begin{equation} \label{eq4}
	E(44 - 55) = \frac{E({\rm H\upbeta - H\upalpha})}{k'({\rm H\upbeta}) - k'({\rm H\upalpha})}
\end{equation}
\begin{equation} \label{eq5}
	\text{Similarly,} \quad E(\uplambda - {\rm H\upbeta}) = [k'(\uplambda) - k'({\rm H\upbeta})] E(44 - 55)
\end{equation} 
\begin{flalign}
	\text{ On the other hand, since\ } I(\uplambda) = F(\uplambda)\ 10^{0.4\,A(\uplambda)}, \text{it follows that}
	\nonumber
\end{flalign}
\begin{flalign} \label{eq6}
	E({\rm H\upbeta - H\upalpha}) &= -2.5~\left\{ {\rm log} \left[ \frac{F({\rm H\upbeta})}{F({\rm H\upalpha})} \right] -
	{\rm log} \left[ \frac{I({\rm H\upbeta})}{I({\rm H\upalpha})} \right] \right\} \nonumber\\
	&= 2.5~{\rm log} \left[ \frac{F({\rm H\upalpha})/F({\rm H\upbeta})}{I({\rm H\upalpha})/I({\rm H\upbeta})} \right]
\end{flalign}
Combining equations \ref{eq3}, \ref{eq4} and \ref{eq6}, we get:
\begin{flalign} \label{eq7}
	{\rm log} \left[ \frac{I(\uplambda)}{I({\rm H\upbeta})} \right] &= {\rm log} \left[ \frac{F(\uplambda)}
	{F({\rm H\upbeta})} \right] \nonumber \\
	&+ \frac{k'(\uplambda) - k'({\rm H\upbeta})}{k'({\rm H\upbeta}) - k'({\rm H\upalpha})} {\rm log}
	\left[ \frac{F({\rm H\upalpha})/F({\rm H\upbeta})}{I({\rm H\upalpha})/I({\rm H\upbeta})} \right]
\end{flalign}

Two more useful quantities need to be evaluated; the logarithmic H$\upbeta$ extinction coefficient,
${\rm c(H\upbeta) = log \left[ \frac{I({\rm H\upbeta})}{F({\rm H\upbeta})} \right]}$,
and the colour excess
$E(B - V)$. ${\rm c(H\upbeta)}$ can be obtained from equations \ref{eq1}, \ref{eq2}, \ref{eq4} and \ref{eq6}:
\begin{equation*} 
	k'({\rm H\upbeta}) = \frac{A({\rm H\upbeta}) - A(44)}{E(44 - 55)} =
	\frac{2.5\ {\rm c(H\upbeta)}}{E(44 - 55)} - R(55)
\end{equation*}
which gives
\begin{flalign} \label{eq8}
	{\rm c(H\upbeta)} &= 0.4\ E(44 - 55)\ [k'({\rm H\upbeta}) + R(55)] \nonumber \\
	&= \frac{k'({\rm H\upbeta}) + R(55)}{k'({\rm H\upbeta}) - k'({\rm H\upalpha})}\ {\rm log}
	\left[ \frac{F({\rm H\upalpha})/F({\rm H\upbeta})}{I({\rm H\upalpha})/I({\rm H\upbeta})} \right]
\end{flalign}
Then, the dereddening relation \ref{eq7} can be written as:
\begin{equation} \label{eq9}
	\frac{I(\uplambda)}{I({\rm H\upbeta})} = \frac{F(\uplambda)}{F({\rm H\upbeta})} 10^{{\rm c(H\upbeta)} f(\uplambda)},
	\text{where} \: f(\uplambda) = \frac{k'(\uplambda) - k'({\rm H\upbeta})}{k'({\rm H\upbeta}) + R(55)}
\end{equation}

We used a series of cubic splines to fit the F19 average extinction curve in $k'(\uplambda)$
for the Milky Way,
which corresponds to $R(V) = 3.1$ and $R(55) = 3.02$, in the wavelength range 3300--10000 \AA.
The anchor points were placed at values of $x$ (inverse wavelength) from Table 3 in F19 and
interpolation on the fitted curve gives $k'({\rm H\upbeta}) = 0.59183$ and 
$k'({\rm H\upalpha}) = -0.57246$.
Then, equation \ref{eq8} becomes:
\begin{equation} \label{eq10}
		{\rm c(H\upbeta)} = 3.1022\ {\rm log} \left[ \frac{F({\rm H\upalpha})/F({\rm H\upbeta})}
		{I({\rm H\upalpha})/I({\rm H\upbeta})} \right]
\end{equation}
The authors in F19  also provide values for the colour excess ratio $r=E(B-V) / E(44-55)$ which, however,
depend on the effective temperature of the observed stars and the overall extinction
[as measured by $E(44 - 55)$].
For each value of $E(44 - 55)$, we have chosen as representative the fixed value that would give
the same area under the curve ($r,~T_{\rm eff}$), that is:
\begin{equation*}
	\langle r\left[ E(44 - 55) \right] \rangle = \frac{\int_{T_{\rm eff,min}}^{T_{\rm eff,max}} r \,dT_{\rm eff}}
	{\int_{T_{\rm eff,min}}^{T_{\rm eff,max}} \,dT_{\rm eff}} \text{, and taking the average over the}
\end{equation*}
two $E(44 - 55)$ values, we obtain
\begin{flalign} \label{eq11}
	E(B- V) &= 0.98\ E(44 - 55) = \frac{0.98\times2.5}{k'({\rm H\upbeta}) + R(55)} \nonumber\\
	&= 0.6873\ {\rm c(H\upbeta)}
\end{flalign}
We give in Table~\ref{tab:ext_func} values of the extinction function $f(\uplambda$) (Eq. \ref{eq9}) for several
optical emission lines of astrophysical interest. 
\begin{table*}
	\caption{Extinction function (defined in Eq.~\ref{eq9}) for several optical emission lines.}
	\label{tab:ext_func}
	\begin{tabular*}{0.98\textwidth}{lclclclclc}
		\hline
		\centering
		$\lambda$(\AA)\hspace{4.5mm}ID & f($\uplambda$) & $\uplambda$(\AA)\hspace{4.5mm}ID & f($\uplambda$)
		& $\uplambda$(\AA)\hspace{4.5mm}ID & f($\uplambda$) & $\uplambda$(\AA)\hspace{4.5mm}ID & f($\uplambda$)
		& $\uplambda$(\AA)\hspace{4.5mm}ID & f($\uplambda$) \\
		\hline
		3703.86\, \ion{H}{I}     & 0.2879 & 4387.93\, \ion{He}{I} & 0.1160 & 6101.83\, [\ion{K}{IV}] & -0.2580
		& 7331.40\, [\ion{Ar}{IV}] & -0.4266 & 8776.97\, \ion{He}{I} & -0.5657 \\
		3705.02\, \ion{He}{I}    & 0.2875 & 4437.55\, \ion{He}{I} & 0.1016 & 6118.20\, \ion{He}{II} & -0.2601
		& 7451.43\, [\ion{Ar}{III}]& -0.4380 & 8799.00\, \ion{He}{II} & -0.5681 \\
		3711.97\, \ion{H}{I}  & 0.2852 & 4471.50\, \ion{He}{I} & 0.0908 & 6233.80\, \ion{He}{II} & -0.2747
		& 7499.84\, \ion{He}{I} & -0.4421 & 8816.64\, \ion{He}{I} & -0.5700 \\
		3721.63\, [\ion{S}{III}] & 0.2822 & 4541.59\, \ion{He}{II} & 0.0693 & 6300.34\, [\ion{O}{I}] & -0.2835
		& 7530.83\, [\ion{Cl}{IV}] & -0.4448 & 8845.37\, \ion{He}{I} & -0.5731 \\
		3721.94\, \ion{H}{I}     & 0.2821 & 4625.53\, [\ion{Ar}{V}] & 0.0478 & 6310.80\, \ion{He}{II} & -0.2850
		& 7592.74\, \ion{He}{II} & -0.4503 & 8914.77\, \ion{He}{I} & -0.5806 \\
		3726.03\, [\ion{O}{II}]  & 0.2809&  4685.68\, \ion{He}{II} & 0.0349 & 6312.10\, [\ion{S}{III}]& -0.2851
		& 7751.06\, [\ion{Ar}{III}] & -0.4685 & 8929.11\, \ion{He}{II} & -0.5822 \\
		3728.82\, [\ion{O}{II}]  & 0.2801 & 4711.37\, [\ion{Ar}{IV}] & 0.0298 & 6363.78\, [\ion{O}{I}] & -0.2924
		& 7751.12\, [\ion{Ar}{III}] & -0.4685 & 8997.02\, \ion{He}{I} & -0.5894 \\
		3734.37\, \ion{H}{I}    & 0.2785 & 4713.17\, \ion{He}{I} & 0.0295 & 6406.30\, \ion{He}{II} & -0.2986
		& 7751.43\, [\ion{Ar}{III}] & -0.4686 & 9068.60\, [\ion{S}{III}] & -0.5968 \\
		3750.15\, \ion{H}{I}     & 0.2741 & 4714.17\, [\ion{Ne}{IV}] & 0.0293 & 6527.11\, \ion{He}{II}& -0.3169
		& 7816.16\, \ion{He}{I} & -0.4775 & 9108.54\, \ion{He}{II} & -0.6009 \\
		3770.63\, \ion{H}{I}     & 0.2684 & 4715.66\, [\ion{Ne}{IV}] & 0.0290 & 6548.10\, [\ion{N}{II}]& -0.3201
		& 8045.63\, [\ion{Cl}{IV}] & -0.5050 & 9123.60\, [\ion{Cl}{II}] & -0.6024 \\
		3797.90\, \ion{H}{I}     & 0.2600 & 4724.15\, [\ion{Ne}{IV}] & 0.0273 & 6562.77\, ${\rm H\upalpha}$
		& -0.3224
		& 8116.60\, \ion{He}{I} & -0.5103 & 9174.49\, \ion{He}{I} & -0.6074 \\
		3805.74\, \ion{He}{I}    & 0.2575 & 4725.62\, [\ion{Ne}{IV}] & 0.0271 & 6583.50\, [\ion{N}{II}] & -0.3256
		& 8168.91\, \ion{He}{I} & -0.5137 & 9210.33\, \ion{He}{I} & -0.6108 \\
		3819.62\, \ion{He}{I}    & 0.2531 & 4740.17\, [\ion{Ar}{IV}] & 0.0243 & 6678.16\, \ion{He}{I} & -0.3400
		& 8236.77\, \ion{He}{II} & -0.5177 & 9303.19\, \ion{He}{I} & -0.6191 \\
		3835.39\, ${\rm H\upeta}$ & 0.2481 & 4861.33\, ${\rm H\upbeta}$ & 0.0000 & 6683.20\, \ion{He}{II}& -0.3408
		& 8237.15\, \ion{He}{I} & -0.5177 & 9344.93\, \ion{He}{II} & -0.6226 \\
		3868.75\, [\ion{Ne}{III}] & 0.2383 & 4921.93\, \ion{He}{I} & -0.0177 & 6716.44\, [\ion{S}{II}]& -0.3458
		& 8265.71\, \ion{He}{I} & -0.5194 & 9367.03\, \ion{He}{II} & -0.6244 \\
		3888.65\, \ion{He}{I}   & 0.2330 & 4931.80\, [\ion{O}{III}] & -0.0209 & 6730.82\, [\ion{S}{II}]& -0.3479
		& 8421.99\, \ion{He}{I} & -0.5306 & 9463.58\, \ion{He}{I} & -0.6317 \\
		3889.05\, ${\rm H\upzeta}$  & 0.2329 & 4958.91\, [\ion{O}{III}] & -0.0297 & 6795.00\, [\ion{K}{IV}]
		& -0.3573 & 8444.69\, \ion{He}{I } & -0.5325 & 9516.63\, \ion{He}{I} & -0.6353 \\
		3926.54\, \ion{He}{I}    & 0.2241 & 5006.84\, [\ion{O}{III}] & -0.0446 & 6890.88\, \ion{He}{I} 
		& -0.3710 & 8451.20\, \ion{He}{I} & -0.5331 & 9530.60\, [\ion{S}{III}] & -0.6361 \\
		3964.73\, \ion{He}{I}    & 0.2162 & 5015.68\, \ion{He}{I} & -0.0471 & 7005.67\, [\ion{Ar}{V}] & -0.3868
		& 8480.85\, [\ion{Cl}{III}] & -0.5357 & 9603.44\, \ion{He}{I} & -0.6404 \\
		3967.46\, [\ion{Ne}{III}] & 0.2157 & 5047.74\, \ion{He}{I} & -0.0555 & 7065.25\, \ion{He}{I} & -0.3948
		& 8486.31\, \ion{He}{I} & -0.5362 & 9625.70\, \ion{He}{I} & -0.6416 \\
		3970.07\, ${\rm H\upepsilon}$ & 0.2151 & 5191.82\, [\ion{Ar}{III}]& -0.0926 & 7135.80\, [\ion{Ar}{III}]
		& -0.4038 & 8519.35\, \ion{He}{II} & -0.5392 & 9702.71\, \ion{He}{I} & -0.6451 \\
		4009.26\, \ion{He}{I}     & 0.2067 & 5197.90\, [\ion{N}{I}] & -0.0943 & 7160.56\, \ion{He}{I} & -0.4069
		& 8528.99\, \ion{He}{I} & -0.5401 & 9762.15\, \ion{He}{II} & -0.6474 \\
		4026.21\, \ion{He}{I}     & 0.2026 & 5200.26\, [\ion{N}{I}] & -0.0950 & 7177.50\, \ion{He}{II} & -0.4090
		& 8566.92\, \ion{He}{II} & -0.5438 & 9824.13\, [\ion{C}{I}] & -0.6492 \\
		4068.60\, [\ion{S}{II}]   & 0.1912 & 5411.52\, \ion{He}{II} & -0.1451 & 7237.26\, [\ion{Ar}{IV}]& -0.4162
		& 8578.69\, [\ion{Cl}{II}] & -0.5450 & 9850.26\, [\ion{C}{I}] & -0.6498 \\
		4076.35\, [\ion{S}{II}]  & 0.1890 & 5517.66\, [\ion{Cl}{III}]& -0.1681 & 7262.76\, [\ion{Ar}{IV}]
		& -0.4191
		& 8581.87\, \ion{He}{I} & -0.5453 & 10023.10\, \ion{He}{I} & -0.6514 \\
		4101.74\, ${\rm H\updelta}$  & 0.1815 & 5537.60\, [\ion{Cl}{III}]& -0.1732 & 7281.35\, \ion{He}{I}& -0.4212
		& 8608.31\, \ion{He}{I} & -0.5479 & 10027.72\, \ion{He}{I} & -0.6514 \\
		4120.84\, \ion{He}{I}     & 0.1758 & 5577.34\, [\ion{O}{I}] & -0.1819 & 7298.04\, \ion{He}{I} & -0.4231
		& 8626.19\, \ion{He}{II} & -0.5498 & 10123.61\, \ion{He}{II} & -0.6503 \\
		4143.76\, \ion{He}{I}     & 0.1692 & 5754.60\, [\ion{N}{II}] & -0.2059 & 7318.92\, [\ion{O}{II}]
		& -0.4253 & 8632.97\, \ion{He}{I} & -0.5505 & 10311.00\, \ion{He}{I} & -0.6442 \\
		4168.97\, \ion{He}{I}     & 0.1625 & 5875.66\, \ion{He}{I} & -0.2240 & 7319.99\, [\ion{O}{II}]
		& -0.4254 & 8648.25\, \ion{He}{I} & -0.5520 & \\
		4340.47\, ${\rm H\upgamma}$  & 0.1268 & 6036.70\, \ion{He}{II} & -0.2490 & 7329.67\, [\ion{O}{II}]
		& -0.4264 & 8733.43\, \ion{He}{I} & -0.5610 & \\
		4363.21\, [\ion{O}{III}]  & 0.1218 & 6074.10\, \ion{He}{II} & -0.2543 & 7330.73\, [\ion{O}{II}]
		& -0.4266 & 8739.97\, \ion{He}{I} & -0.5617 & \\
		\hline
	\end{tabular*}
\end{table*}

\bsp	
\label{lastpage}
\end{document}